\documentclass[%
aip,
jcp,%
amsmath,amssymb,
reprint,%
]{revtex4-1}
\usepackage[latin1]{inputenc}
\usepackage{amsmath}
\usepackage{amsfonts}
\usepackage{amssymb}
\usepackage{graphicx}
\usepackage{physics}
\usepackage{color}
\usepackage{chngcntr}

\newcommand{\pP}{\ensuremath{\mathcal{P}}}
\newcommand{\pQ}{\ensuremath{\mathcal{Q}}}
\newcommand{\pL}{\ensuremath{\mathcal{L}}}
\newcommand{\pK}{\ensuremath{\mathcal{K}}}

\newcommand{\sys}{\ensuremath{\mathrm{s}}}
\newcommand{\nuc}{\ensuremath{\mathrm{n}}}
\newcommand{\el}{\ensuremath{\mathrm{e}}}
\newcommand{\eq}{\ensuremath{\mathrm{eq}}}
\newcommand{\sing}{\ensuremath{\mathrm{S}}}
\newcommand{\trip}{\ensuremath{\mathrm{T}}}
\newcommand{\ipi}{\ensuremath{\mathrm{I}}}
\renewcommand{\op}[1]{\ensuremath{{#1}}}
\newcommand{\nint}[1]{\ensuremath{{#1}^\nuc}}

\begin{document}

\title{Spin-selective electron transfer reactions of radical pairs: beyond the Haberkorn master equation}
\author{Thomas P. Fay}
\email{thomas.fay@ccc.ox.ac.uk}
\affiliation{Department of Chemistry, University of Oxford, Physical and Theoretical Chemistry Laboratory, South Parks Road, Oxford, OX1 3QZ, UK}
\author{Lachlan P. Lindoy}
\affiliation{Department of Chemistry, University of Oxford, Physical and Theoretical Chemistry Laboratory, South Parks Road, Oxford, OX1 3QZ, UK}
\author{David E. Manolopoulos}
\affiliation{Department of Chemistry, University of Oxford, Physical and Theoretical Chemistry Laboratory, South Parks Road, Oxford, OX1 3QZ, UK}

\begin{abstract}
	Radical pair recombination reactions are normally described using a quantum mechanical master equation for the electronic and nuclear spin density operator. The electron spin state selective (singlet and triplet) recombination processes are described with a Haberkorn reaction term in this master equation. Here we consider a general spin state selective electron transfer reaction of a radical pair and use Nakajima-Zwanzig theory to derive the master equation for the spin density operator, thereby elucidating the relationship between non-adiabatic reaction rate theory and the Haberkorn reaction term. A second order perturbation theory treatment of the diabatic coupling naturally results in the Haberkorn master equation with an additional reactive scalar electron spin coupling term. This term has been neglected in previous spin chemistry calculations, but we show that it will often be quite significant. We also show that beyond second order in perturbation theory, i.e., beyond the Fermi golden rule limit, an additional reactive singlet-triplet dephasing term appears in the master equation. A closed form expression for the reactive scalar electron spin coupling in terms of the Marcus theory parameters that determine the singlet and triplet recombination rates is presented. By performing simulations of radical pair reactions with the exact Hierarchical Equations of Motion (HEOM) method, we demonstrate that our master equations provide a very accurate description of radical pairs undergoing spin-selective non-adiabatic electron transfer reactions. The existence of a reactive electron spin coupling may well have implications for biologically relevant radical pair reactions such as those which have been suggested to play a role in avian magnetoreception.
\end{abstract}

\maketitle
	
\section{Introduction}
The radical pair mechanism has been used extensively to describe magnetic field effects in many chemical  reactions.\cite{Brocklehurst2002,Steiner1989,Rodgers2009} In these reactions the key intermediate is the radical pair. This intermediate state undergoes spin state selective reactions -- the reaction product and reaction rate depend on the spin state of the electrons in the radical pair. If the singlet and triplet electron spin states are close in energy, they can coherently interconvert due to weak magnetic interactions in the radicals, such as hyperfine interactions with nuclear spins. The coherent spin dynamics and spin-selective reaction pathways can give rise to large magnetic field effects on the dynamics and quantum yields of these reactions.\cite{Brocklehurst2002,Steiner1989,Rodgers2009,Wasielewski2006}

Radical pair reactions are conventionally described using the reduced density operator for the spin degrees of freedom of the radical pair, $\rho_{\rm s}(t)$. The unitary evolution of this density operator due to the interactions contained in the spin Hamiltonian $H_{\rm s}$ is given by the usual Liouville--von Neumann equation. The non-unitary reactive dynamics are then conventionally treated by adding an additional term to this equation, known as the Haberkorn term.\cite{Haberkorn1976,Johnson1970,Evans1973} Overall, the full master equation for the spin density operator is
\begin{align}\label{trad-me-eqn}
\dv{t}\rho_{\rm s}(t) = -\frac{i}{\hbar}\left[H_{\rm s},\rho_{\rm s}(t)\right]-\left\{K_{\rm s},\rho_{\rm s}(t)\right\},
\end{align}
where $[\cdot, \cdot]$ is a commutator and $\{\cdot,\cdot\}$ is an anti-commutator. The Haberkorn reaction operator is
\begin{align}
K_{\rm s} = \frac{k_\mathrm{S}}{2}P_\sing+\frac{k_\mathrm{T}}{2}P_\trip,
\end{align}
in which $P_\sing$ and $P_\trip$ are the projection operators onto singlet and triplet electronic states of the radical pair and $k_\mathrm{S}$ and $k_\mathrm{T}$ are the singlet and triplet recombination rate constants. 
\begin{figure}
	\includegraphics[width=0.325\textwidth]{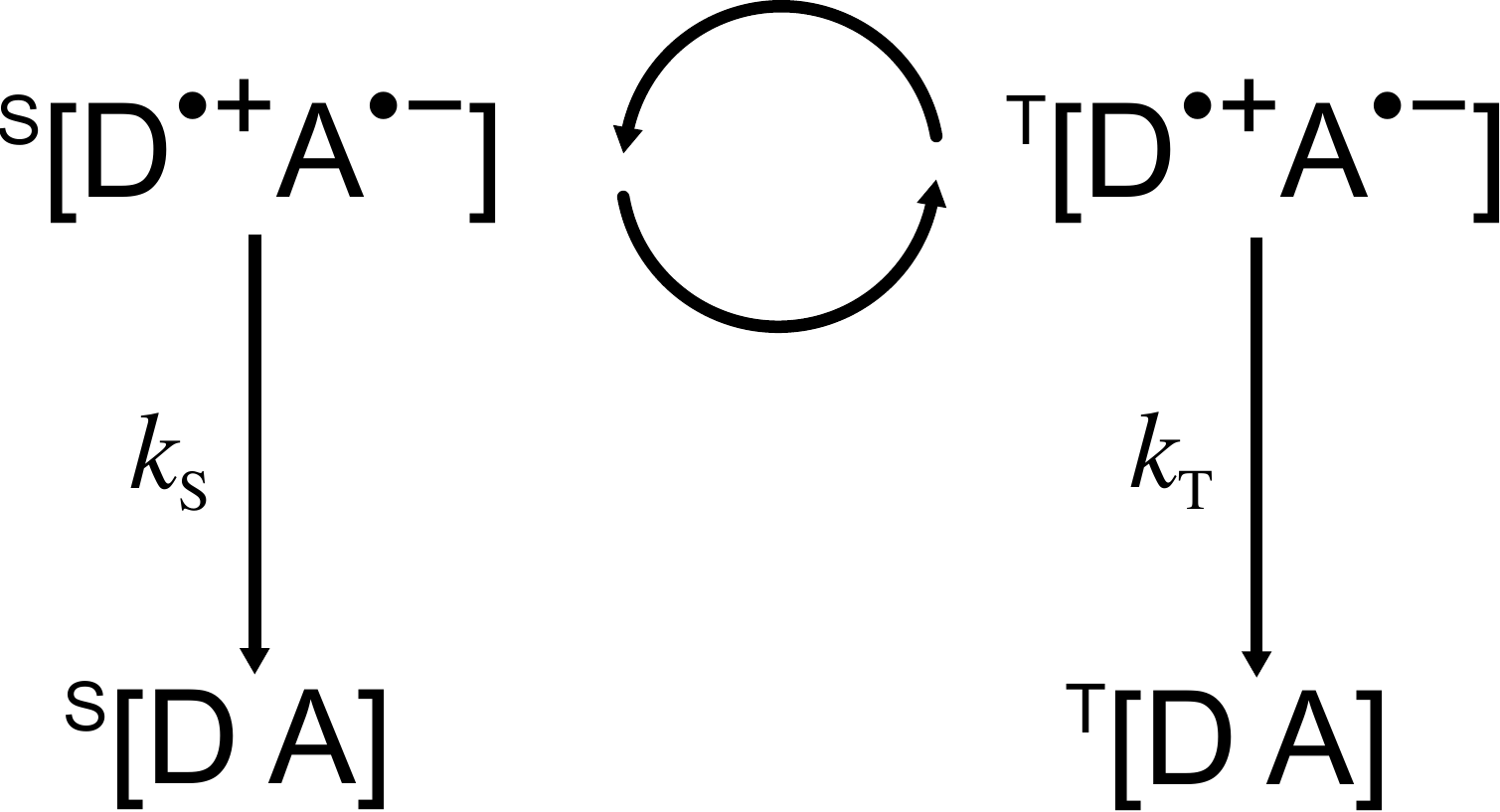}
	\caption{The radical pair mechanism for a photo-generated donor (D) acceptor (A) radical pair system. }\label{rp-scheme}
\end{figure}

This form of master equation has been used successfully for over 40 years to explain magnetic field effects on radical pair reactions. However, in recent years several alternative master equations have been suggested,\cite{Jones2010,Jones2011,Kominis2009} leading to some debate in the spin chemistry literature as to which master equation correctly describes the radical pair mechanism.\cite{Kominis2009,Jones2010,Jones2011,Purtov2010,Ivanov2010,Tiersch2012,Clausen2014} Alternative approaches based on quantum measurement theory have been proposed such as the Jones-Hore\cite{Jones2010,Jones2011} master equation, which is the same as the Haberkorn master equation but includes an additional singlet-triplet dephasing term of the form
\begin{align}\label{jh-term-eqn}
-\frac{k_\sing + k_\trip}{2}\left[P_\sing \rho_{\rm s}(t)P_\trip + P_\trip \rho_{\rm s}(t)P_\sing\right].
\end{align}
Other more complex master equations based on quantum measurement theory have also been suggested.\cite{Kominis2009} 

The Haberkorn master equation predicts that singlet-triplet coherences in the spin density operator should decay at a rate of $(k_\sing+k_\trip)/2$, whereas quantum measurement based master equations predict larger decay rates of coherences. Maeda \textit{et al.}\cite{Maeda2013} used this distinction to experimentally test the validity of the various master equations for a carotene-porphorin-fullerene triad radical pair, and found that the singlet-triplet coherence decay rate of this radical pair was uniquely consistent with the Haberkorn master equation.

It is therefore somewhat surprising that a general derivation of the Haberkorn master equation from chemical reaction rate theory has not been presented so far in the literature. A derivation starting from a microscopic description of the electron pair recombination reaction was originally alluded to by Evans \textit{et al.} in 1973,\cite{Evans1973} and eventually presented by Ivanov \textit{et al.} in 2010.\cite{Ivanov2010} However, their derivation was based on a highly simplified model of the radical pair reaction. The nuclear degrees of freedom were treated as a harmonic bath linearly coupled to the radical pair and product states, and the total density operator was assumed to remain in the form $W(t) = \rho_{\rm s}(t)\rho_{\rm n}^{\rm eq}$, where $\rho_{\rm n}^{\rm eq}$ is the equilibrium density operator of the nuclear motion bath. These assumptions rarely hold for real radical pairs, which have anharmonic radical pair and product states with different equilibrium geometries, leading to significant coupling between the electronic and nuclear evolution.

In order to establish a more rigorous connection between chemical reaction rate theory and the Haberkorn master equation, we shall consider an important subset of radical pair reactions -- non-adiabatic electron transfers in radical ion pairs.\cite{Wasielewski2006,Lewis2014} Typically these systems consist of an electron donor, D, and an electron acceptor, A. The system is first energetically excited, often by absorption of a photon. The excited state undergoes an electron transfer to generate a $[\text{D}^{\bullet +}\text{A}^{\bullet -}]$ radical ion pair. This radical ion pair then undergoes coherent interconversion between its singlet and triplet states and spin-selective electron transfers to singlet and triplet product states, as illustrated schematically in Fig. \ref{rp-scheme}.

We shall present a derivation of the Haberkorn master equation for these spin selective electron transfers of radical pairs based on the well-established theory of non-adiabatic electron transfer reactions.\cite{Sparpaglione1988,Evans1995,Golosov2001} In section \ref{model-sec} we describe the diabatic state model for spin selective radical pair electron transfer reactions. We outline the general theory and approximations used to derive master equations for this model in section \ref{gen-qme-sec}, and in section \ref{qme-sec} we derive explicit master equations for electron transfer reactions of radical pairs. In section \ref{tests-sec} we perform exact simulations for a set of model radical pair systems, explicitly including all nuclear degrees of freedom, and compare the results to those of our master equations and the Haberkorn master equation. In section \ref{disc-sec} we discuss the significance of our results and suggest some experiments that might be performed to verify them.

\section{Non-Adiabatic Reactions of Radical Pairs}\label{model-sec}

Many experimentally examined radical pair systems undergo spin state selective electron transfer reactions.\cite{Rodgers2009,Steiner1989} In electron transfer reactions there is a breakdown of the Born-Oppenheimer approximation and there are non-adiabatic transitions between different Born-Oppenheimer (adiabatic) potential energy surfaces. One can also describe these reactions using diabatic potential energy surfaces.\cite{Nitzan2006,VanVoorhis2010} In the non-adiabatic limit, it is the off-diagonal coupling between diabatic states that gives rise to electron transfer. 

\begin{figure*}
	\includegraphics[width=0.7\textwidth]{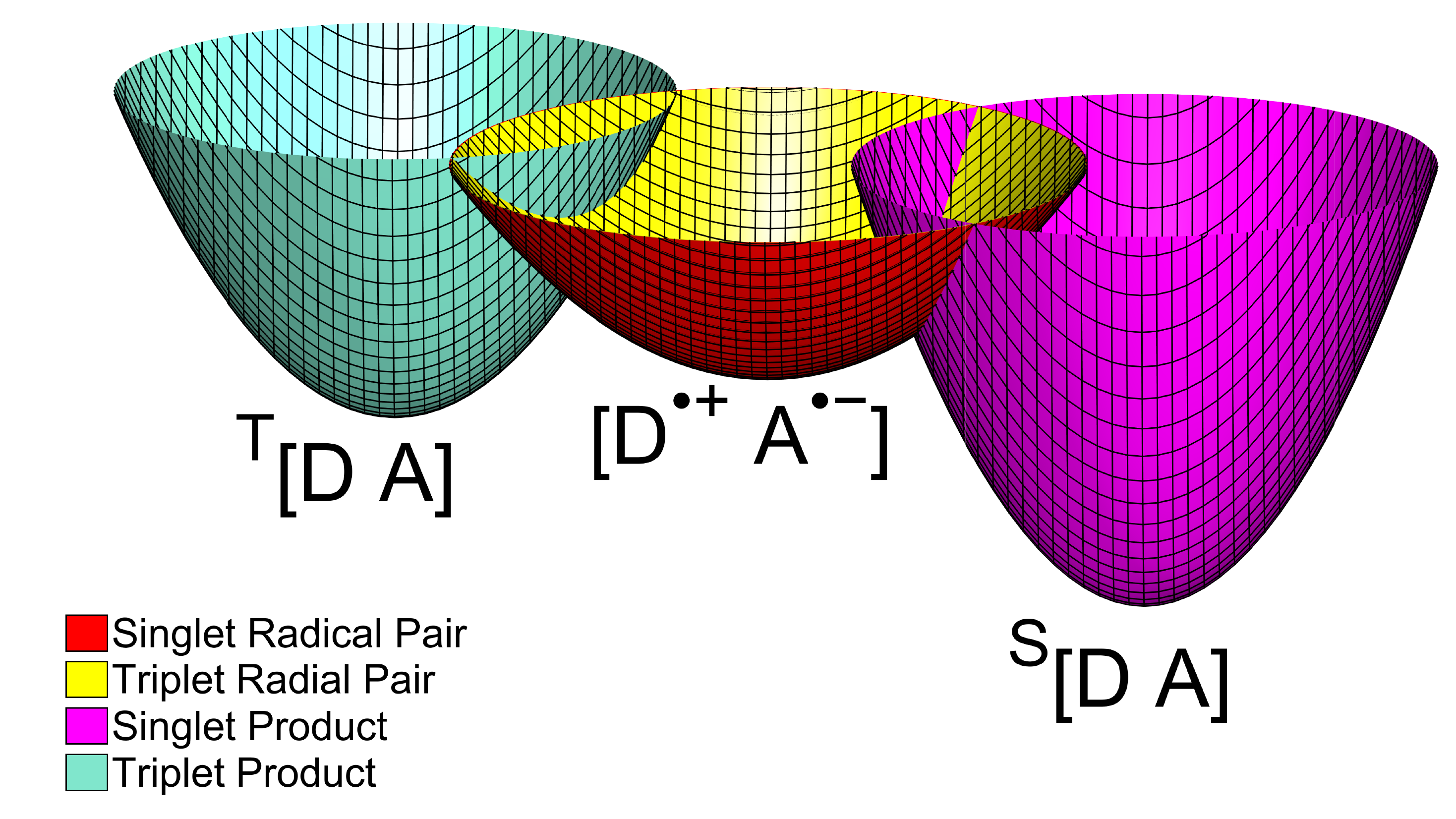}
	\caption{A schematic diabatic potential energy diagram for a radical pair system with recombinative singlet and triplet electron transfer pathways. The singlet and triplet radical pair diabats are very close in energy whereas the singlet and triplet product surfaces have a very different structure. \label{diabat-diag}}
\end{figure*}
In our approach we consider two sets of diabatic electronic states -- the radical pair states $\ket{1}\ket{\sing}$ and $\ket{1}\ket{\trip_m}$, the singlet product state $\ket{2}\ket{\sing}$ and the triplet product states $\ket{2}\ket{\trip_m}$. Conservation of spin in the electron transfers means there exists a coupling only between radical pair states and product states with the same spin state. A schematic representation of the problem is illustrated in Fig. \ref{diabat-diag}. The Hamiltonian for the full radical pair system, including all spin, nuclear\footnote{In this paper we will use the term ``nuclear'' to refer only to the spatial degrees of freedom of the nuclei and the term ``spin'' will refer generally to nuclear and electronic spin degrees of freedom.} and electronic degrees of freedom, is
\begin{align}
\begin{split}
H =\  &H_1\dyad{1} + H_2\dyad{2} \\ 
&+ \Delta_\sing P_\sing\left( f_\sing\dyad{1}{2}+f_\sing^\dagger\dyad{2}{1}\right) \\
&+ \Delta_\trip P_\trip\left( f_\trip\dyad{1}{2}+f_\trip^\dagger\dyad{2}{1}\right),
\end{split}\label{hamiltonian-eqn}
\end{align}
where $H_j$ is the Hamiltonian for the nuclear and spin degrees of freedom in electronic state $j$. The third term contains the diabatic coupling between the singlet radical pair state and the singlet product state, and the fourth term is the same but for the triplet radical pair and product states. $\Delta_\sing$ and $\Delta_\trip$ are the diabatic coupling constants for the singlet surfaces and triplet surfaces respectively. $f_\sing$ and $f_\trip$ are operators on the nuclear degrees of freedom but in the following discussion we will make the Condon approximation, in which $f_\sing$ and $f_\trip$ are assumed to be independent of the nuclear coordinates and replaced with identity operators.\cite{Nitzan2006} 

The radical pair Hamiltonian, $H_1$, may be divided into three terms: a spin 
term, $H_{1\sys}$, a nuclear term, $H_{1\nuc}$, and a nuclear-spin coupling term, $H_{1\nuc\sys}$,
\begin{align}
H_1 = H_{1\sys} + H_{1\nuc} + H_{1\nuc\sys}.
\end{align}
The product Hamiltonian, $H_2$, consists of the nuclear term for each of the spin states $H_{2\nuc}^\sing$ and $H_{2\nuc}^\trip$, accompanied by appropriate spin-state projection operators
\begin{align}
H_2 = P_\sing H_{2\nuc}^{\sing} + P_\trip H_{2\nuc}^{\trip} .
\end{align}
We assume that there is no coupling between singlet and triplet product states. This model for the radical pair system is simply the multi-state generalisation of the standard model of electron transfer.\cite{Sparpaglione1988} Thus far we have made no assumptions about the forms of the different diabatic potential energy surfaces, only that the Hamiltonian can be separated as presented. 

The coupling between the radical pair spin states is contained in the spin Hamiltonians $H_{1\sys}$ and $H_{1\nuc\sys}$. We shall assume that the radical pair singlet and triplet potential energy surfaces lie very close in energy (i.e., that the radical pair has a small exchange coupling), in which case the spin dynamics will be much slower than the nuclear dynamics. The nuclear-spin coupling term $H_{1\nuc\sys}$ causes spin relaxation of the radical pair.\cite{Goldman2001} In the following discussion we assume the spin relaxation is very slow and so we ignore this term in $H_1$. From this point on we also set $\Delta_\trip=0$, which means we ignore the triplet recombination pathway. This is done to simplify the discussion and notation, but the generalisation to $\Delta_\trip\neq 0$ is straightforward and is presented in appendix \ref{trippath-app}. 

The ensemble of radical ion pairs is described by the density operator $W(t)$ for the full system. We assume that initially there are no coherences between electronic states and that the nuclear degrees of freedom are in thermal equilibrium on each diabatic potential energy surface. Therefore the density operator starts in a state of the form
\begin{align}
W(0) = \rho_{1\sys}(0)\rho_{1\nuc}^\eq\dyad{1} + \rho_{2\sys}(0)\rho_{2\nuc}^\eq\dyad{2} .
\end{align}
The initial spin density operators are denoted by $\rho_{j\sys}(0)$. We are ignoring the triplet reaction, and therefore for state 2 this initial spin density operator should only contain singlet components and thus should satisfy
\begin{align}
\rho_{2\sys}(0) = P_\sing \rho_{2\sys}(0) P_\sing.
\end{align}
$\rho_{j\nuc}^\eq$ is the thermal equilibrium density operator for the nuclear degrees of freedom on diabat $j$,
\begin{align}
\rho_{j\nuc}^\eq = \frac{1}{Z_{j\nuc}}e^{-\beta H_{j\nuc}},
\end{align}
where $\beta^{-1}=k_\mathrm{B} T$ and $Z_{j\nuc} = \Tr_\nuc[e^{-\beta H_{j\nuc}}]$. Because we are ignoring the triplet product we have that $H_{2\nuc}\equiv H_{2\nuc}^\sing$. The radical pair and product potential energy surfaces are in general very different so $\rho_{1\nuc}^\eq\neq\rho_{2\nuc}^\eq$. This is important and was overlooked in the approach taken by Ivanov \textit{et al.} in Ref. \onlinecite{Ivanov2010}.

Previous derivations of master equations for radical pair reactions have only considered simple system-bath models with harmonic baths and linear system-bath couplings.\cite{Ivanov2010, Kominis2009} We would like to emphasise that the present approach is more general -- the diabatic potential energy surfaces may be highly anharmonic and the results presented extend straightforwardly to the case of more complex coupling between the electronic states [i.e., for general $f_\sing$ and $f_\trip$ in Eq.~(4)].

\section{Quantum Master Equations}\label{gen-qme-sec}

The Liouville--von Neumann equation describes the dynamics of the full density operator of the system,
\begin{align}
\dv{t}W(t) = \mathcal{L}W(t).
\end{align}
The Liouvillian superoperator $\mathcal{L}$ is defined by
\begin{align}
\mathcal{L}A = -\frac{i}{\hbar}\left[H,A\right],
\end{align} 
for any operator $A$ on the Hilbert space. The expectation value of an operator, $O$, is given by
\begin{align}
\ev{O} = \Tr[O W(t)],
\end{align}
where $\Tr$ denotes the trace over the full Hilbert space. 

Exact evolution of the density operator for the full system, including all spin, electronic and nuclear degrees of freedom, is a formidable task given the large size of the full Liouville space (the space of operators on the Hilbert space). However, in spin chemistry we are rarely interested in the dynamics of the full system. More often we are interested only in the populations of the radical pair electron spin states and the corresponding product states. This information is fully contained in the reduced density operators for the spin degrees of freedom of the radical pair and product states,
\begin{subequations}
\begin{align}
\rho_{1\sys}(t) &= \Tr_\nuc[\ev{W(t)}{1}], \\
\rho_{2\sys}(t) &= \Tr_\nuc[\ev{P_\sing W(t)P_{\sing}}{2}].
\end{align}
\end{subequations}
Here $\Tr_\nuc$ denotes the partial trace over the nuclear degrees of freedom. The aim of this work is therefore to obtain a set of equations for the dynamics of the reduced density operators -- these equations are referred to as master equations. 

\subsection{Liouville Space Projection Superoperators}

In order to obtain master equations for the reduced density operators it is useful to introduce Liouville space projection superoperators. These project operators in the full Liouville space to some subspace of Liouville space.\cite{Nakajima1958,Zwanzig1960} We require that our projection superoperator, $\mathcal{P}$, has the following property
\begin{subequations}\label{projred-eqn}
	\begin{align}
	\rho_{1\sys}(t) &= \Tr_\nuc[\ev{\mathcal{P}W(t)}{1}], \\
	\rho_{2\sys}(t) &= \Tr_\nuc[\ev{P_\sing (\mathcal{P}W(t))P_\sing}{2}].
	\end{align}
\end{subequations}
If we can obtain a master equation for the projected density operator, $\mathcal{P}W(t)$, then from this we can straightforwardly obtain the equations of motion for the reduced density operators. The master equation for $\mathcal{P}W(t)$ is also simplified if our initial density operator is fully contained within the projected subspace,
\begin{align}
\pP W(0) = W(0). \label{projinit-eqn}
\end{align}
As such, we need to define a projection superoperator $\pP$ satisfying these properties. To this end, we define $\mathcal{P}$ as a sum of two other projection superoperators,
\begin{align}
\pP = \pP_1 +  \pP_2.
\end{align}
These projection superoperators are defined as follows,
\begin{subequations}\label{projdefs-eqn}
	\begin{align}
	\pP_1 A &= \rho_{1\nuc}^\eq \dyad{1} \Tr_\nuc[\ev{A}{1}], \\
	\pP_2 A &= \rho_{2\nuc}^\eq \dyad{2} \Tr_\nuc[\ev{P_\sing A P_\sing}{2}]  ,
	\end{align}
\end{subequations}
where $A$ is any Hilbert space operator. We see that $\pP_j^2 = \pP_j$ so these are indeed projection superoperators. Also $\pP_1\pP_2=\pP_2\pP_1=0$ and therefore $\pP$ is also a projection superoperator. Noting that the projected density operator is related to the reduced density operators by
\begin{align}
\pP_j W(t)  = \rho_{j\nuc}^\eq \dyad{j}\rho_{j\sys}(t),
\end{align}
we see that $\pP$ also clearly satisfies properties \eqref{projred-eqn} and \eqref{projinit-eqn}.

\subsection{The Nakajima-Zwanzig Equation}

To derive an exact equation of motion for $\mathcal{P}W(t)$, we divide the Hamiltonian into a reference part $H_0$ and a perturbation $V$ as
\begin{align}
H = H_0 + V.
\end{align}
For the Hamiltonian in Eq.~\eqref{hamiltonian-eqn} we define the reference Hamiltonian $H_0$ and the perturbation $V$ as
\begin{subequations}
	\begin{align}
		H_0 &= (H_{1\nuc}+H_{1\sys})\dyad{1} + H_2 \dyad{2}, \\
		V &=  \Delta \left(\dyad{1}{2}+\dyad{2}{1}\right) P_\sing,
	\end{align}
\end{subequations}
where we have explicitly set $\Delta_\trip=0$ and $\Delta_\sing\equiv \Delta$. We have also neglected the spin-nuclear coupling term. Given this we can write the Liouvillian as 
\begin{align}
\mathcal{L} = \mathcal{L}_0 + \mathcal{L}_V,
\end{align}
with $\mathcal{L}_0$ and $\mathcal{L}_V$ defined by
\begin{subequations}
	\begin{align}
	\mathcal{L}_0 A &= -\frac{i}{\hbar}\left[H_0, A\right],\\
	\mathcal{L}_V A &= -\frac{i}{\hbar}\left[V, A\right],
	\end{align}
\end{subequations}
for any operator $A$.  Our Liouville space projection superoperator, $\mathcal{P}$, commutes with the reference Liouvillian,
\begin{align}
	\mathcal{P}\mathcal{L}_0 &= \mathcal{L}_0\mathcal{P}, \label{projpropa-eqn} 
\end{align}
The interaction picture Liouvillian of $V$, $\pL_V^\ipi(t)$, is defined as
\begin{align}
{\pL}_V^\ipi(t)A =e^{-\pL_0 t}\pL_V e^{\pL_0 t}A= -\frac{i}{\hbar}\left[V^{\ipi}(t),A\right], \label{ipiliouvillian-eqn}
\end{align}
where $A$ is any Hilbert space operator, and the interaction picture perturbation operator $V^{\ipi}(t)$ is
\begin{align}
V^{\ipi}(t) = e^{iH_0t/\hbar}Ve^{-iH_0t/\hbar}.
\end{align}
A product of an odd number of interaction picture Liouvillians of $V$ has the following property,
\begin{align}
	\mathcal{P}\mathcal{L}_V^\ipi(t_{2n+1})\cdots\mathcal{L}_V^\ipi(t_1)\mathcal{P} &= 0, \label{projpropb-eqn}
\end{align}
the proof of which is given in appendix \ref{projprop-app}.

Using standard projection superoperator techniques,\cite{Nakajima1958,Zwanzig1960} the equation of motion for the projected density operator is found to be
\begin{align}
\dv{t}\mathcal{P}W(t) &= \mathcal{L}_0\mathcal{P}W(t)+\int_0^t\mathcal{K}(t-t_0)\mathcal{P}W(t_0)\dd{t_0}. \label{nz-eqn}
\end{align}
This is the Nakajima-Zwanzig equation.\cite{Nakajima1958, Zwanzig1960} The kernel $\mathcal{K}(t)$ is given by\cite{Breuer2001}
\begin{align}
\begin{split}
\mathcal{K}(t) =\  &e^{\pL_0t} \pP {\pL}_V^\ipi(t) \\ &\times \left(\mathsf{T}\exp[\int_0^t\pQ {\pL}_V^\ipi(\tau)\dd{\tau}]\right)\pQ\pL_V\pP. \label{kernel-eqn}
\end{split}
\end{align}
Here $\pQ$ is the complementary projection superoperator $\pQ = 1-\pP$, and $\mathsf{T}$ is the chronological time-ordering operator for Liouville space superoperators. We would like to emphasise that Eq.~\eqref{nz-eqn} is formally exact. However it is also no easier to solve than the Liouville--von Neumann equation for the full system. The reason we introduce it is that Eq.~\eqref{nz-eqn} provides a useful starting point for obtaining approximate master equations for the projected density operator. 

\subsection{Incoherent Recombination Approximation}

We are mostly concerned with systems for which the non-adiabatic reaction rate is well defined. For this to be true there must be a separation of time scales between the dynamics of $\pP W(t)$ and that of the kernel $\pK(t)$. This is true if the spin dynamics are much slower than the nuclear dynamics. To formalise this we start by taking the one-sided Fourier transform of the Nakajima-Zwanzig equation. This transform is defined as\cite{Sparpaglione1988}
\begin{align}
\hat{f}(\omega) = \lim_{\eta\rightarrow 0^+}\int_0^\infty e^{+i(\omega+i\eta) t}f(t)\dd{t},
\end{align}
and the inverse transform for $t\geq 0$ is given by
\begin{align}
f(t) = \frac{1}{2\pi}\int_{-\infty}^{\infty}e^{-i\omega t}\hat{f}(\omega)\dd{\omega}.
\end{align}
The transform of Eq.~\eqref{nz-eqn} is
\begin{align}
-i\omega \pP \hat{W}(\omega)-\pP W(0) = \pL_0\pP \hat{W}(\omega)+\hat{\pK}(\omega) \pP \hat{W}(\omega).
\end{align}
If the kernel $\pK(t)$ decays on a much faster time scale than the dynamics of $\pP W(t)$, then $\pP W(\omega)$ will be much more sharply peaked around $\omega = 0$ than $\hat{\pK}(\omega)$. This means that we can approximate $\hat{\pK}(\omega) \pP \hat{W}(\omega)$ as $\hat{\pK}(0) \pP \hat{W}(\omega)$.\cite{Sparpaglione1988} With this approximation and inverting the one-sided Fourier transform we obtain the following Markovian and time homogeneous equation for the projected density operator,
\begin{align}
\dv{t}\pP W(t) = \pL_0\pP W(t)+\hat{\pK}(0) \pP W(t),
\end{align}
in which the superoperator $\hat{\pK}(0)$ is given by
\begin{align}
\hat{\pK}(0) = \int_0^\infty\pK(t_0)\dd{t_0}.
\end{align}
If we cannot make this approximation then a rate constant based description of the reaction is not appropriate for describing the radical pair recombination.

We note that in replacing $\hat{\pK}(\omega)$ with $\hat{\pK}(0)$ we do not affect the long time limit of $\pP W(t)$. This can be seen by taking the formal solution to the one-sided Fourier transform of the master equation,
\begin{align}
\pP \hat{W}(\omega) = -\left(i\omega + \pL_0 + \hat{\pK}(\omega)\right)^{-1} \pP W(0),
\end{align}
and noting that the long-time limit of $\pP W(t)$ is given by 
\begin{align}
\lim_{t\to \infty} \pP W(t) = \lim_{\omega\to 0} (-i\omega)\pP W(\omega).
\end{align}
Clearly replacing $\hat{\pK}(\omega)$ with $\hat{\pK}(0)$ does not affect the long-time limit of $\pP W(t)$. 

\subsection{Field Independent Rate Approximation}

The evaluation of $\hat{\pK}(0)$ is complicated by the $H_{1\sys}$ term appearing in $H_0$. However, if the spin dynamics of the radical pair are much slower than the nuclear dynamics (i.e. if the energy scale of $H_{1\sys}$ is much smaller than that of $H_{j\nuc}$) then we can set $H_{1\sys}$ to zero inside this kernel. The \textit{a posteriori} justification for this is that this approximation gives recombination rate constants that are independent of an applied magnetic field, as is observed experimentally. 

To formalise this, we define $\pK_\nuc(t)=\pK(t)|_{H_{1\sys}=0}$ as the field independent rate kernel,
\begin{align}
\begin{split}
\mathcal{K}_\nuc(t) =\  & \pP {\pL}_V^\nuc(t) \\ &\times \left(\mathsf{T}\exp[\int_0^t\pQ {\pL}_V^\nuc(\tau)\dd{\tau}]\right)\pQ\pL_V\pP, \label{kerneln-eqn}
\end{split}
\end{align}
where we define the nuclear interaction picture Liouvillian as
\begin{align}
{\pL}_V^\nuc(t)A = e^{-\pL_\nuc t} \pL_V e^{\pL_\nuc t} A = -\frac{i}{\hbar}[V^\nuc(t),A],\label{nintliouvillian-eqn}
\end{align}
and the nuclear interaction picture perturbation as
\begin{align}
V^\nuc(t) = e^{i H_\nuc t/\hbar} Ve^{-iH_\nuc t/\hbar},
\end{align}
in which $H_\nuc = H_{1\nuc}\dyad{1} + H_{2\nuc}\dyad{2}P_\sing$. With this approximation the master equation for $\pP W(t)$ is
\begin{align}
\dv{t}\pP W(t) = \pL_\sys\pP W(t) + \pK \pP W(t), \label{master-eqn}
\end{align}
in which the spin Liouvillian is defined as
\begin{align}
	\pL_\sys A = -\frac{i}{\hbar}[H_{1\sys}\dyad{1}{1},A],
\end{align}
and the rate superoperator $\pK$ is defined as
\begin{align}
\pK = \int_0^\infty \pK_\nuc(t_0)\dd{t_0}.
\end{align}
To obtain Eq.~\eqref{master-eqn} we have used the fact that $\pL_0\pP = \pL_\sys\pP$. The key difference between this master equation and others widely used in non-adiabatic reaction rate theory is that $\pK$ is a superoperator on the spin degrees of freedom as well as the electronic and nuclear degrees of freedom. The rate superoperator is still not significantly easier to evaluate than the full kernel $\pK(t)$, so to proceed further we need to use perturbation theory to approximate $\pK$. 

\subsection{Perturbative Expansion}\label{pert-sec}

In the non-adiabatic limit we assume that the diabatic coupling $\Delta$ is small. In this limit we can make a perturbative expansion of the rate kernel.\cite{Sparpaglione1988} This is achieved by expanding the time-ordered exponential in equation \eqref{kernel-eqn}. Property \eqref{projpropb-eqn} of $\pP$ means that all odd terms in the expansion vanish and the rate superoperator can be written as
\begin{align}
\pK = \sum_{k=1}^\infty\pK^{(2k)},
\end{align}
where $\pK^{(2k)}$ is proportional to $\Delta^{2k}$.\cite{Sparpaglione1988} Truncation of this expansion at $\pK^{(2n)}$ gives an approximate master equation with leading order error of $\mathcal{O}(\Delta^{2n+2})$. 

\section{Master Equations For Non-Adiabatic Reactions of Radical Pairs}\label{qme-sec}

Having described the general framework for obtaining perturbative master equations for the dynamics of the projected density operator, we will now explicitly obtain master equations accurate to second and fourth order in the diabatic coupling $\Delta$ for our radical pair model.

Before proceeding, we note that since $\pP = \pP_1+\pP_2$ we can rewrite Eq.~\eqref{master-eqn} as
\begin{align}
\dv{t} \pP_j W(t) = \pL_\sys \pP_jW(t)+\sum_{k=1}^2 \pK_{jk} \pP_k W(t), \label{components-master-eqn}
\end{align}
where
\begin{align}
\pK_{jk} = \pP_j \pK \pP_k.
\end{align}
$\pK_{11}$ is the superoperator describing the loss of the radical pair, and $\pK_{12}$ describes the back reaction process, transferring population from state 2 to 1. Similarly $\pK_{22}$ describes loss from state 2 via the back reaction, and $\pK_{21}$ describes transfer of population from state 1 to state 2 in the forward reaction. This gives a convenient way to separate the terms in the master equations for the reduced density operators.

At this point it is also useful to introduce the following further notation. The perturbation $V$ may be written as 
\begin{align}
V = \Delta(\sigma_+ + \sigma_-)P_\sing,
\end{align}
where $\sigma_+ = \dyad{2}{1}$ and $\sigma_- = \dyad{1}{2}$. We note that only alternating sequences of $\sigma_+$ and $\sigma_-$ are non-zero and that only even alternating sequences connect $\ket{j}$ with itself. For example
\begin{align}
\ev{\sigma_-\sigma_+\cdots\sigma_-\sigma_+}{1}=1,\label{sigmapropa-eqn}
\end{align}
and only odd alternating sequences connect $\ket{1}$ with $\ket{2}$,
\begin{align}
\mel{1}{\sigma_-\sigma_+\cdots\sigma_-\sigma_+\sigma_-}{2}=1.\label{sigmapropb-eqn}
\end{align}
Similarly, $\nint{V}(t)$ in Eq.~\eqref{nintliouvillian-eqn} can be written in terms of $\nint{\sigma}_{\pm}(t)$ as 
\begin{align}
\nint{V}(t) = \Delta\left(\nint{\sigma}_+(t) + \nint{\sigma}_-(t) \right)P_\sing,
\end{align}
where $\nint{\sigma}_+(t)$ is given by
\begin{align}
\nint{\sigma}_+(t) &= \sigma_+ e^{+i H_{2\nuc}t/\hbar}e^{-i H_{1\nuc}t/\hbar},\label{nintsigmapm-eqn}
\end{align}
and $\nint{\sigma}_-(t)=\nint{\sigma}_+(t)^\dag$. These observations allow us to dramatically simplify the multiple commutators appearing in $\pK_{jk}$, as any terms not of these forms vanish.

\subsection{Second Order Master Equation}

The second order term in $\pK$ is given by
\begin{align}
\pK^{(2)} = \int_0^\infty\pP \nint{\pL}_V(t_0)\pL_V\pP\dd{t_0}.
\end{align}
This is obtained by expanding the time ordered exponential in Eq.~\eqref{kerneln-eqn}, retaining only the leading $\mathcal{O}(\Delta^0)$ term. 
First we will use this to evaluate $\pK_{11}^{(2)}$, the second order term appearing in $\pK_{11}$. We can write the Liouvillians in terms of commutators as
\begin{align}
\pP_1 \nint{\pL}_V(t_0)\pL_V\pP_1 W(t) =-\frac{1}{\hbar^2} \pP_1 \left[\nint{V}(t_0),\left[V,\pP_1 W(t) \right]\right].
\end{align}
Writing $\nint{V}(t_0)$ in terms of $\nint{\sigma}_{\pm}(t_0)$ and expanding the commutators, there are 16 terms in this expression for $\pK_{11}^{(2)}$. Using properties \eqref{sigmapropa-eqn} and \eqref{sigmapropb-eqn} of $\sigma_{\pm}$ we can eliminate all but two of these terms, which leaves
\begin{widetext}
\begin{align}
\pK_{11}^{(2)}\pP_1 W(t) = -\frac{\Delta^2}{\hbar^2}\int_0^\infty \dd{t_0}\pP_1 \bigg(\nint{\sigma}_-(t_0)\sigma_+P_\sing(\pP_1 W(t))
+(\pP_1 W(t))P_\sing\sigma_-\nint{\sigma}_+(t_0)\bigg).
\end{align}
With some further manipulations using Eq.~\eqref{projdefs-eqn} and Eq.~\eqref{nintsigmapm-eqn}, we can simplify this to
\begin{align}
\pK_{11}^{(2)}\pP_1 W(t) = -\frac{\Delta^2}{\hbar^2}\int_0^\infty \dd{t_0} \bigg(c^{(2)}_1(t_0)P_\sing(\pP_1 W(t))
+c^{(2)}_1(t_0)^*(\pP_1 W(t))P_\sing\bigg),
\end{align}
\end{widetext}
where the function $c_1^{(2)}(t)$ is defined as
\begin{align}\label{corrfunc1-eqn}
c_1^{(2)}(t) = \Tr_\nuc\left[\rho_{1\nuc}^\eq e^{+iH_{1\nuc}t/\hbar}e^{-iH_{2\nuc}t/\hbar}\right].
\end{align}
The integral of this function from $t=0$ to $\infty$ has both real and imaginary parts. Splitting these parts up we find that the $\pK_{11}^{(2)}$ term can be written as
\begin{align}
\begin{split}
\pK_{11}^{(2)}\pP_1 W(t) = &-\left\{ \frac{k_\mathrm{f}^{(2)}}{2} P_\sing, \pP_1 W(t) \right\} \\ &-\frac{i}{\hbar}\left[ 2J^{(2)} P_\sing, \pP_1 W(t)\right],
\end{split}
\end{align}
where $k_\mathrm{f}^{(2)}$ is the Fermi golden rule non-adiabatic rate constant for the forward reaction,\cite{Wolynes1987}
\begin{align}
k_\mathrm{f}^{(2)} = \frac{2\Delta^2}{\hbar^2} \int_0^\infty \Re\left[c_1^{(2)}(t)\right]\dd{t} \label{kf-2-eqn}
\end{align}
and $J^{(2)}$ is a reactive contribution to the electron spin coupling given by
\begin{align}
J^{(2)} = \frac{\Delta^2}{2\hbar}\int_{0}^\infty \Im\left[c_1^{(2)}(t)\right]\dd{t}.\label{coup-2-eqn}
\end{align}
We therefore see that the $\pK_{11}^{(2)}$ term contains a Haberkorn reaction term in which the rate constant is the well-known Fermi golden rule non-adiabatic electron transfer rate. However it also contains a reactive contribution to the scalar electron spin coupling, which is not present in the traditional Haberkorn treatment. We will demonstrate later that this term is not in general negligible for radical pair reactions.

We follow the same procedure to evaluate the $\pK_{12}^{(2)}$ term. First expanding the double commutator as before, writing $\nint{V}(t_0)$ in terms of $\nint{\sigma}_{\pm}(t_0)$ and using the properties of $\sigma_{\pm}$, we obtain
\begin{widetext}
	\begin{align}
	\pK_{12}^{(2)}\pP_2 W(t) = \frac{\Delta^2}{\hbar^2}\int_0^\infty \dd{t_0}\pP_1 \bigg(\nint{\sigma}_-(t_0)(\pP_2 W(t))\sigma_+
	+\sigma_-(\pP_1 W(t))\nint{\sigma}_+(t_0)\bigg).
	\end{align}
\end{widetext}
Again with some manipulations using Eq.~\eqref{projdefs-eqn} and Eq.~\eqref{nintsigmapm-eqn}, we can simplify this to
\begin{align}
	\pK_{12}^{(2)}\pP_2 W(t) = k_\mathrm{b}^{(2)}\mathcal{S}_-\pP_2 W(t).
\end{align}
In this expression $k_\mathrm{b}^{(2)}$ is the Fermi golden rule rate constant for the back reaction,
\begin{align}
k_\mathrm{b}^{(2)} = \frac{2\Delta^2}{\hbar^2} \int_0^\infty \Re\left[c_2^{(2)}(t)\right]\dd{t},
\end{align}
where $c_2^{(2)}(t)$ is given by
\begin{align}\label{corrfunc2-eqn}
c_2^{(2)}(t) = \Tr_\nuc\left[\rho_{2\nuc}^\eq e^{+iH_{2\nuc}t/\hbar}e^{-iH_{1\nuc}t/\hbar}\right],
\end{align}
and $\mathcal{S}_-$ is a superoperator that transfers a projected operator $\pP_2 A$ from the projected subspace of $\pP_2$ to the projected subspace of $\pP_1$,
\begin{align}
\mathcal{S}_-\pP_2 A = \rho_{1\nuc}^\eq\dyad{1}\Tr_\nuc\left[\ev{\pP_2 A}{2}\right].
\end{align}
Repeating these steps for $\pK_{21}^{(2)}$ and $\pK_{22}^{(2)}$ we find
\begin{align}
\pK_{21}^{(2)}\pP_1 W(t) = k_\mathrm{f}^{(2)}\mathcal{S}_+\pP_1 W(t),
\end{align}
and 
\begin{align}
\pK_{22}^{(2)}\pP_2 W(t) = -k_\mathrm{b}^{(2)}\pP_2 W(t),
\end{align}
where we have defined $\mathcal{S}_+$ as the superoperator that transfers the singlet component of a projected operator $\pP_1 A$ to the projected subspace of $\pP_2$,
\begin{align}
\mathcal{S}_+ \pP_1 A = \rho_{2\nuc}^\eq\dyad{2}P_\sing\Tr_\nuc\left[\ev{\pP_1 A}{1}\right]P_\sing.
\end{align}

Combining these results we now have a full master equation for $\pP W(t)$. By tracing out the nuclear and electronic degrees of freedom as in Eq.~\eqref{projred-eqn}, and using $P_\sing = \frac{1}{4}-\vb{S}_1\cdot\vb{S}_2$ (where we use unitless spin operators), we obtain the following set of master equations for the reduced spin density operators for the two electronic states,
\begin{widetext}
\begin{subequations}\label{meq-second-order-eqn}
	\begin{gather}
	\dv{t}\rho_{1\sys}(t) = -\frac{i}{\hbar}\left[H_{1\sys},\rho_{1\sys}(t)\right] -\frac{i}{\hbar}\left[(-2J^{(2)})\vb{S}_1\cdot\vb{S}_2,\rho_{1\sys}(t)\right]  - \left\{ \frac{k_\mathrm{f}^{(2)}}{2} P_\sing, \rho_{1\sys}(t) \right\} + k_\mathrm{b}^{(2)} P_\sing \rho_{2\sys}(t) P_\sing, \label{meq-rp-second-order-eqn}\\
	\dv{t}\rho_{2\sys}(t) = k_\mathrm{f}^{(2)}P_\sing \rho_{1\sys}(t)P_\sing - k_\mathrm{b}^{(2)} \rho_{2\sys}(t). \label{meq-prod-second-order-eqn}
	\end{gather}
\end{subequations}
\end{widetext}
These quantum master equations have leading order error of $\mathcal{O}(\Delta^4)$ within the incoherent recombination approximation. The first term in the master equation for the radical pair reduced density operator $\rho_{1\sys}(t)$ is the normal coherent spin evolution term and the second term is an additional reactive contribution to the scalar electron spin coupling. The third term in \eqref{meq-rp-second-order-eqn} is a Haberkorn type term describing the singlet-selective reaction and the fourth term describes the back reaction. The master equation for the singlet product state, equation \eqref{meq-prod-second-order-eqn}, is a simple first order kinetic equation, with the first term describing the forward reaction and the second term describing the back reaction. 

Equations \eqref{meq-rp-second-order-eqn} and \eqref{meq-prod-second-order-eqn} are a key result of this paper. The Haberkorn master equation for the radical pair spin-density operator has been derived from a first principles description of the radical pair reaction. This first principles approach also naturally results in a reactive contribution to the scalar electron spin coupling, which emerges as a correction to conventional Haberkorn master equation. In appendix \ref{trippath-app} we give the full version of these quantum master equations including the triplet recombination pathway. One important difference in this case is that the contribution to $J^{(2)}$ from the triplet reaction pathway has the opposite sign because $P_\trip = 3/4 + \vb{S}_1\cdot\vb{S}_2$.

We may understand the origin of the reactive electron-spin coupling as follows. The coupling between two states is known to cause shifts in the energies of the states and the lowest order correction to the energy levels is of order $\Delta^2$. Because in the model we have considered thus far, the diabatic coupling only exists between singlet radical pair and product states, only the singlet state of the radical pair is shifted in energy. The net result of this is an electron spin coupling which is related to the thermally averaged energy shift of radical pair singlet state. When the triplet recombination pathway is included as in appendix B, a net reactive exchange coupling emerges from the difference between the thermally averaged energy shifts of the singlet and triplet radical pair states.

\subsection{Fourth Order Master Equation}

As demonstrated above, a second order treatment of the non-adiabatic coupling naturally yields the Haberkorn master equation for the radical pair reaction with an additional electron spin coupling term. We now go beyond the non-adiabatic limit and examine the fourth order contribution in $\Delta$ to the master equation. We will see that a fourth order treatment of the diabatic coupling not only gives the expected fourth order corrections to the rate constants and electron spin coupling, but also gives rise to a reactive singlet-triplet dephasing term in the master equation, similar to that introduced by Jones and Hore\cite{Jones2010,Jones2011}.

The full expression for $\pK^{(4)}$ is
\begin{align}
\begin{split}
\pK^{(4)} = \int_0^\infty&\dd{t_0}\int_0^{t_0}\dd{t_1}\int_0^{t_1}\dd{t_2}\\ &\times\pP \nint{\pL}_V(t_0) \nint{\pL}_V(t_1)(1-\pP)\nint{\pL}_V(t_2)\pL_V\pP.
\end{split}
\end{align}
As before we consider each component $\pK_{jk}^{(4)}$ of $\pK^{(4)}$ in turn. 

For $\pK_{11}^{(4)}$ we can write terms in the integrand as nested commutators, for example
\begin{align}
\begin{split}
\pP_1 &\nint{\pL}_V(t_0) \nint{\pL}_V(t_1)\nint{\pL}_V(t_2)\pL_V\pP_1 W(t) = \\
&\frac{1}{\hbar^4}\pP_1 \left[\nint{V}(t_0),\left[\nint{V}(t_1),\left[\nint{V}(t_2),\left[V,\pP_1 W(t)\right]\right] \right]\right].
\end{split}
\end{align}
We then expand the nested commutators and write $\nint{\op{V}}(t_n)$ in terms of $\nint{\op{\sigma}}_\pm(t_n)$. Overall there are 768 terms appearing in this expansion, but using the properties of $\sigma_{\pm}$ we can eliminate the majority of these terms. This leads to a Haberkorn term and an electron coupling term as in $\pK_{11}^{(2)}$. Additionally, non-vanishing terms with $P_\sing$ on both sides of $\pP_1 W(t)$ appear, for example terms of the form
\begin{align}
\pP_1\left(\sigma_-\sigma_+P_\sing(\pP_1 W(t))P_\sing\sigma_-\sigma_+\right).
\end{align}
Noting that $\op{P}_\sing = 1-\op{P}_\trip$, we can write $P_\sing (\pP_1 W(t)) P_\sing$ as 
\begin{align}
\begin{split}
P_\sing (\pP_1 &W(t)) P_\sing = \frac{1}{2}\left\{P_\sing,(\pP_1 W(t))\right\} \\
&- \frac{1}{2}\left(P_\trip(\pP_1 W(t))P_\sing + P_\sing(\pP_1 W(t))P_\trip  \right).
\end{split}
\end{align}
We notice that this produces terms in $\pK_{11}^{(4)}$ which contribute to the Haberkorn reaction term and an additional singlet-triplet dephasing term as in Eq.~\eqref{jh-term-eqn}. After some manipulations, the final result for $\pK_{11}^{(4)}$ is
\begin{widetext}
\begin{align}
\pK_{11}^{(4)}\pP_1 \op{W}(t) = -\left\{ \frac{k_\mathrm{f}^{(4)}}{2}\op{P}_\sing, \pP_1 \op{W}(t) \right\} -\frac{i}{\hbar}\left[2J^{(4)}\op{P}_\sing,\pP_1 \op{W}(t)\right] -k^{(4)}_\mathrm{d} \bigg(\op{P}_\sing(\pP_1 W(t))\op{P}_\trip+\op{P}_\trip(\pP_1 W(t))\op{P}_\sing\bigg).
\end{align}
\end{widetext}
Here we define $k_\mathrm{f}^{(4)}$ as the fourth order contribution to the forward rate constant, $J^{(4)}$ as the fourth order contribution to the reactive electron spin coupling and $k_\mathrm{d}^{(4)}$ as the fourth order singlet-triplet dephasing rate. The explicit expressions for these quantities are somewhat lengthy, involving triple time integrals, and are given in appendix \ref{fourthorder-app}. Repeating this for the other components of the fourth order rate superoperator we find
\begin{align}
\pK_{12}^{(4)} \pP_2 \op{W}(t) &= k_\mathrm{b}^{(4)}\mathcal{S}_-\pP_2 \op{W}(t), \\
\pK_{21}^{(4)} \pP_1 \op{W}(t) &= k_\mathrm{f}^{(4)}\mathcal{S}_+\pP_1 \op{W}(t), \\
\pK_{22}^{(4)} \pP_2 \op{W}(t) &= -k_\mathrm{b}^{(4)}\pP_2\op{W}(t),
\end{align}
where $k_\mathrm{b}^{(4)}$ is the fourth order contribution to the back reaction rate. 

Combining these expressions for the fourth order superoperator components with the second order terms and Eq.~\eqref{components-master-eqn} and taking the trace over the nuclear and electronic state degrees of freedom as in Eq.~\eqref{projred-eqn}, we find that the fourth order master equations for the reduced density operators for the spin degrees of freedom are
\begin{widetext}
	\begin{subequations}\label{meq-fourth-order-eqn}
		\begin{gather}
		\dv{t}\rho_{1\sys}(t) = -\frac{i}{\hbar}\left[H_{1\sys},\rho_{1\sys}(t)\right] -\frac{i}{\hbar}\left[(-2J)\vb{S}_1\cdot\vb{S}_2,\rho_{1\sys}(t)\right]  - \left\{ \frac{k_\mathrm{f}}{2} P_\sing, \rho_{1\sys}(t) \right\}  -k_\mathrm{d}\bigg(P_\sing\rho_{1\sys}(t)P_\trip+P_\trip\rho_{1\sys}(t)P_\sing\bigg) + k_\mathrm{b} P_\sing \rho_{2\sys}(t) P_\sing \label{meq-rp-fourth-order-eqn}\\
		\dv{t}\rho_{2\sys}(t) = k_\mathrm{f}P_\sing \rho_{1\sys}(t)P_\sing - k_\mathrm{b} \rho_{2\sys}(t). \label{meq-prod-fourth-order-eqn}
		\end{gather}
	\end{subequations}
\end{widetext}
Here we define $J = J^{(2)}+J^{(4)}$, $k_\mathrm{f} = k_\mathrm{f}^{(2)}+k_\mathrm{f}^{(4)}$, $k_\mathrm{b} = k_\mathrm{b}^{(2)}+k_\mathrm{b}^{(4)}$ and $k_\mathrm{d} = k_\mathrm{d}^{(4)}$. These quantum master equations have leading order error, within the incoherent recombination and field-independent rate approximations, of $\mathcal{O}(\Delta^6)$. The rate constants $k_\mathrm{f}$ and $k_\mathrm{b}$ are the same rate constants as those appearing in other formal expressions for the non-adiabatic rate to fourth order.\cite{Sparpaglione1988,Golosov2001} Equations \eqref{meq-rp-fourth-order-eqn} and \eqref{meq-prod-fourth-order-eqn} are the second key result of this paper. They show that fourth order contributions in the diabatic coupling to the recombination dynamics give rise to a singlet-triplet dephasing term in the master equation for the radical pair spin density operator in addition to the Haberkorn term and a reactive electron spin coupling. 

The physical origin of the dephasing can be understood as follows. Fourth and higher order terms in $\Delta$ in the kernel contain transition state recrossing contributions.\cite{Cho1997,Evans1995,Coalson1994,Golosov2001} These recrossing contributions project the radical pair spin system onto the singlet state, which is known to result in singlet-triplet dephasing.\cite{Kattnig2016} This results in an additional decay of coherences as well as a reduction in the total recombination rate constant.

We can see that higher order terms in $\Delta$ appearing in $\pK$ will not introduce additional spin superoperators to this equation because $\op{P}_\sing^2 = \op{P}_\sing$, but the parameters $k_\mathrm{f}$, $k_\mathrm{b}$, $J$ and $k_\mathrm{d}$ will all contain higher order contributions. Therefore the master equation accurate to \textit{all} orders in $\Delta$ has the form of the fourth order master equation, Eq.~\eqref{meq-fourth-order-eqn}. This is of the same form of the Jones and Hore master equation\cite{Jones2010,Jones2011}, but unlike in the Jones-Hore master equation the dephasing rate $k_\mathrm{d}$ is not necessarily equal to $k_\mathrm{f}/2$. We should note that the dephasing only appears at fourth order in $\Delta$ so the leading order terms in $\Delta$ in the master equation are the Haberkorn term and the reactive electron spin coupling. In the supplementary information we show that this is true when the triplet recombination pathway is included as well.

\subsection{Marcus-Hush Theory Limit}

Marcus-Hush theory provides an approximate formula for the rate of an electron transfer reaction in the non-adiabatic (second order in $\Delta$) and classical limits.\cite{Marcus1956, Hush1958, Zwanzig2001} Here we extend the Marcus theory for the rate constant to obtain an analogous expression for the reactive contribution to the scalar electron spin coupling. 

One way to derive Marcus theory is to start from the spin-boson model\cite{Caldeira1983} for the electron transfer.\cite{Zwanzig2001} Within this model the $N$ nuclear degrees of freedom are treated as a bath of harmonic modes and $H_{1\nuc}$ and $H_{2\nuc}$ can be written as
\begin{subequations}\label{spinboson-eqn}
\begin{align}
H_{1\nuc} &= \sum_{k=1}^N \left(\frac{P_k^2}{2m_k} + \frac{1}{2}m_k\omega_k^2 Q_k^2 + c_k Q_k\right), \\
H_{2\nuc} &= \sum_{k=1}^N \left(\frac{P_k^2}{2m_k} + \frac{1}{2}m_k\omega_k^2 Q_k^2 - c_k Q_k\right)-\epsilon,
\end{align}
\end{subequations}
where $P_k$ and $Q_k$ are the momentum and position operators for bath mode $k$, and $m_k$ and $\omega_k$ are the mass and angular frequency of the mode. $\epsilon$ is the bias which for this model is exactly the negative of the free energy difference between the states, $\Delta_\mathrm{r}G$. The reorganisation energy, $\lambda$, is related to $\omega_k$, $m_k$ and  the coupling constants $c_k$ by
\begin{align}
\lambda = \sum_{k=1}^N\frac{2 c_k^2}{m_k \omega_k^2}.
\end{align}
Within the classical Marcus-Hush approximation we replace all quantum mechanical operators with the corresponding classical variables, i.e. $P_k\rightarrow p_k$ and $Q_k\rightarrow q_k$, and we replace the trace over bath mode $k$ in Eq.~\eqref{corrfunc1-eqn} with
\begin{align}
\Tr_k \rightarrow \frac{1}{2\pi\hbar}\int_{-\infty}^\infty\dd{p_k}\int_{-\infty}^\infty\dd{q_k}.
\end{align}
Integrating out the momenta, the classical approximation to $c_1^{(2)}(t)$ is then 
\begin{align}
\begin{split}
c_{1,\mathrm{cl}}^{(2)}(t) =&e^{{\beta \lambda}/{4}+i{\epsilon t}/{\hbar}} \prod_{k=1}^N\left(\frac{2\pi k_\mathrm{B}T}{m_k\omega_k^2}\right)^{\frac{1}{2}}\\
&\times \int_{-\infty}^\infty \dd{q_k}  e^{-{(m_k \omega_k^2 q_k^2 + 2c_k q_k)}/{2k_\mathrm{B}T} +2ic_kq_k t/\hbar}.
\end{split}
\end{align}
Performing the integrals and evaluating the product, we find that this gives
\begin{align}
c_{1,\mathrm{cl}}^{(2)}(t) = e^{i(\epsilon -\lambda) t/\hbar - k_\mathrm{B}T \lambda (t/\hbar)^2},\label{class-c2-eqn}
\end{align}
and using equation \eqref{kf-2-eqn} we obtain
\begin{align}
k_{\mathrm{f,cl}}^{(2)} = \frac{\Delta^2}{\hbar}\sqrt{\frac{\pi }{k_\mathrm{B}T\lambda}}e^{-{(\lambda-\epsilon)^2}/{4\lambda k_\mathrm{B}T}},\label{marcus-rate-eqn}
\end{align}
which is the well-known Marcus-Hush theory expression for the electron-transfer rate. Now using Eq.~\eqref{class-c2-eqn} in equation \eqref{coup-2-eqn} we obtain the following expression for the reactive electron spin coupling,
\begin{align}
J_{\mathrm{cl}}^{(2)} = \frac{\Delta^2}{4}\sqrt{\frac{\pi }{k_\mathrm{B}T\lambda}}e^{-{(\lambda-\epsilon)^2}/{4\lambda k_\mathrm{B}T}}\mathrm{erfi}\left(\frac{\epsilon-\lambda}{2\sqrt{k_\mathrm{B}T\lambda}}\right),\label{marcus-spincoup-eqn}
\end{align}
where $\mathrm{erfi}(x)$ is the imaginary error function,
\begin{align}
\mathrm{erfi}(x) = \frac{2}{\sqrt{\pi}}\int_0^x e^{z^2}\dd{z}.
\end{align}
This is the third main result of this paper, a formula for the reactive scalar electron spin coupling in the non-adiabatic limit which depends only on parameters in the Marcus-Hush formula for the electron transfer rate. It should be noted that the integral of $c_1^{(2)}(t)$ appearing in Eqs. \eqref{kf-2-eqn} and \eqref{coup-2-eqn} is an analytic function of $\epsilon$ in the upper half-plane and therefore the forward rate constant, $k_\mathrm{f}^{(2)}$, and the reactive electron spin coupling, $J^{(2)}$, are related by a Kramers-Kronig relation. It can also be shown that this is true for anharmonic diabatic surfaces.

\begin{figure}[t]
	\includegraphics[width=0.48\textwidth]{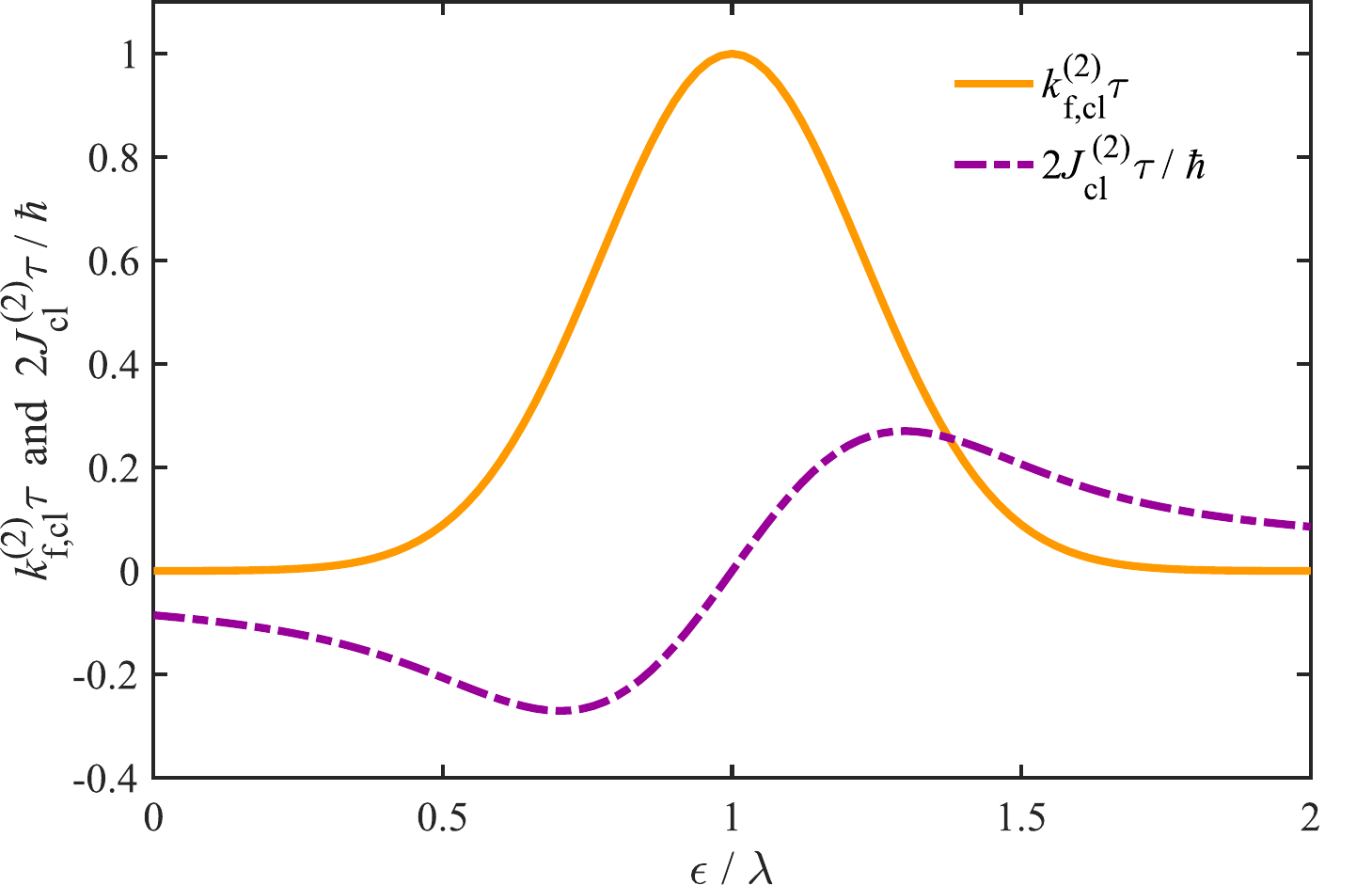}
	\caption{The Marcus-Hush theory rate constant and the reactive scalar electron spin coupling as a function of $\epsilon/\lambda$ for $\lambda$ = 1 eV and $T$ = 300 K. Here $\tau$ is defined by $\tau^{-1} = {\Delta^2}\sqrt{{\pi }/({\hbar^2 k_\mathrm{B}T\lambda}})$.}\label{marcus-fig}
\end{figure}

The Marcus-Hush theory values for $k_\mathrm{f,cl}^{(2)}$ and $2J_{\mathrm{cl}}^{(2)}/\hbar$ are plotted in Fig. \ref{marcus-fig} for $T=300\ \mathrm{K}$ and $\lambda=1\ \mathrm{eV}$. We see that in both the symmetric electron transfer and strongly inverted regimes, the scalar electron spin coupling can in fact be orders of magnitude larger than the rate constant. For example, for this set of parameters when $\epsilon=0$ or $\epsilon=2\lambda$, the ratio of these quantities, $|2J_{\mathrm{cl}}^{(2)}/\hbar k_\mathrm{f,cl}^{(2)}|$ is over $1000$. However, close to the maximum in the Marcus-Hush theory rate, the rate constant is much larger than the reactive contribution to the scalar electron spin coupling, i.e. $k_\mathrm{f,cl}^{(2)}\gg|2J_{\mathrm{cl}}^{(2)}/\hbar|$. Clearly  in general we cannot say whether or not the reactive contribution to the scalar electron spin coupling is negligible. 

\section{Numerical Tests}\label{tests-sec}

In deriving the master equations for the radial pair electron transfer reaction we introduced three approximations: 1) the incoherent recombination approximation, 2) the field independent rate approximation and 3) the perturbative expansion of the rate superoperator. In order to demonstrate that these approximations are valid for radical pair electron transfers, we now compare master equation results to an exactly soluble model for a condensed phase electron transfer reaction.

\subsection{Model Radical Pair Systems}

In general the potential energy surfaces of radical pair systems are highly complex and solving the Liouville--von Neumann equation for the full radical pair for arbitrary potential energy surfaces is generally not tractable. However, for small systems and for certain potential energy surfaces methods exist that can produce numerically exact results for the quantum dynamics. We consider one such exactly soluble system -- the spin-boson model.\cite{Caldeira1983, heom2} In this model the electronic system is coupled to an infinite bath of harmonic oscillators. The spin-boson model has been applied extensively as a model of condensed phase electron transfer.\cite{Berkelbach2012,Zwanzig2001,Richardson2015,Wang2003,Cho1997,Mavros2014}  The Hamiltonians for the radical pair and product surfaces are given by Eq.~\eqref{spinboson-eqn}. The coupling constants $c_k$ are related to the spectral density for the system $J(\omega)$ by
\begin{align}
J(\omega) = \frac{\pi}{2}\sum_{k=1}^N \frac{c_k^2}{m_k\omega_k}\delta(\omega-\omega_k).
\end{align}
In our model we use the Debye spectral density, which is given by
\begin{align}
J(\omega) = \frac{\lambda}{2}\frac{\omega \omega_c}{\omega^2 +  \omega_c^2},
\end{align}
in which $\lambda$ is the reorganisation energy and $\omega_c$ is the cut-off frequency. For the spin-boson model we can obtain exact results using the Hierarchical Equations of Motion method.\cite{heom2}

We consider two model radical pair spin Hamiltonians and initial conditions, in which we only include the singlet reaction pathway. In Model I we take the spin Hamiltonian $H_{1\sys}$ to be $0$ and we consider a system initially in a superposition of singlet and triplet states $(\ket{\sing}+\ket{\trip_0})/\sqrt{2}$ so the initial spin density operator is
\begin{align}
\rho_{1\sys}(0) = \frac{1}{2}\left(\dyad{\sing}+\dyad{\sing}{\trip_0} + \dyad{\trip_0}{\sing} + \dyad{\trip_0}\right),
\end{align}
and we choose $\rho_{2\sys}(0)$ to be $0$. This somewhat artificial model is chosen to allow easy comparison of the decay rates of the populations and the singlet-triplet coherences as a function of the diabatic coupling $\Delta$. For this we can exactly calculate all parameters appearing in the second and fourth order master equations using the analytic expressions given in appendices \ref{fourthorder-app} and \ref{spinboson-app}.

In Model II we consider a radical pair in an external magnetic field, $B$, with a single $I=1/2$ nuclear spin in one of the radicals. The spin Hamiltonian for Model II is
\begin{align}
H_{1\sys} = \omega_0 S_{1z} + \omega_0 S_{2z} + a \vb{I} \cdot \vb{S}_1.
\end{align}
$\omega_0 = - \hbar \gamma_\el B  $ is the Zeeman frequency of the electron spin, with gyromagnetic ratio $\gamma_\el$ and $S_{iz}$ is the $z$ component of the unitless electron spin operator for radical $i$. The hyperfine coupling constant between the nuclear and electron spins in radical 1 is denoted by $a$, and $\vb{S}_1$ and $\vb{I}$ are the electron and nuclear spin vector operators for radical 1. The initial state is chosen to be pure singlet radical pair so $\rho_{1\sys}(0)$ is given by 
\begin{align}
\rho_{1\sys}(0) = \frac{1}{2}P_\sing,
\end{align}
and $\rho_{2\sys}(0)=0$. This is chosen as a realistic model of a singlet-born photogenerated radical ion pair, in particular to demonstrate the importance of including the reactive contribution to the scalar electron spin coupling and the validity of the field independent rate approximation.

\subsection{Simulation Details}

\subsubsection{Model Parameters}

In both models we use the Debye spectral density with  $\hbar \omega_c = 1.24\text{ meV} $. This relatively small value for $\hbar\omega_c$ is used to increase the computational efficiency of the exact calculations. In Model I we set the bias $\epsilon$ between the radical pair state and singlet product state to $0$ eV and the reorganisation energy $\lambda = 0.25$ eV and in Model II we set $\epsilon = 0.1$ eV and $\lambda  = 0.5$ eV. All simulations are run at a temperature of $T=300$ K. We vary the diabatic coupling $\Delta$ between 0.1 and 3 meV for Model I and in Model II we set $\Delta = 0.1 $ meV. The spin system parameters in Model II are chosen to be $\omega_0 / | \hbar \gamma_\el|= 0.5$ mT and $a / | \hbar \gamma_\el| = 1.5$ mT as typical radical parameters.

\subsubsection{Exact Simulations}

Numerically exact results for the spin-boson model are obtained using the well-established Hierarchical Equations of Motion (HEOM) method.\cite{heom1, heom2}
We use a Matsubara expansion of the bath correlation functions and a frequency based truncation scheme\cite{heom3} to construct the hierarchy of auxiliary density operators required for this method.  
Using this truncation scheme, results converged to graphical accuracy for Model I and Model II are obtained using a maximum frequency of 250 $\omega_c$.  
This corresponds to a hierarchy of 369 auxiliary density operators, including contributions from the first two Matsubara modes.

In order to efficiently integrate over the long timescales required here, it is necessary to use an adaptive order and time step Taylor series integrator. The scaling scheme of Shi \textit{et al.}\cite{heom4} is used in order to allow for effective control of the errors in the integrator.

\subsubsection{Master Equations}
\begin{figure*}[t]
	\includegraphics[width=1.0\textwidth]{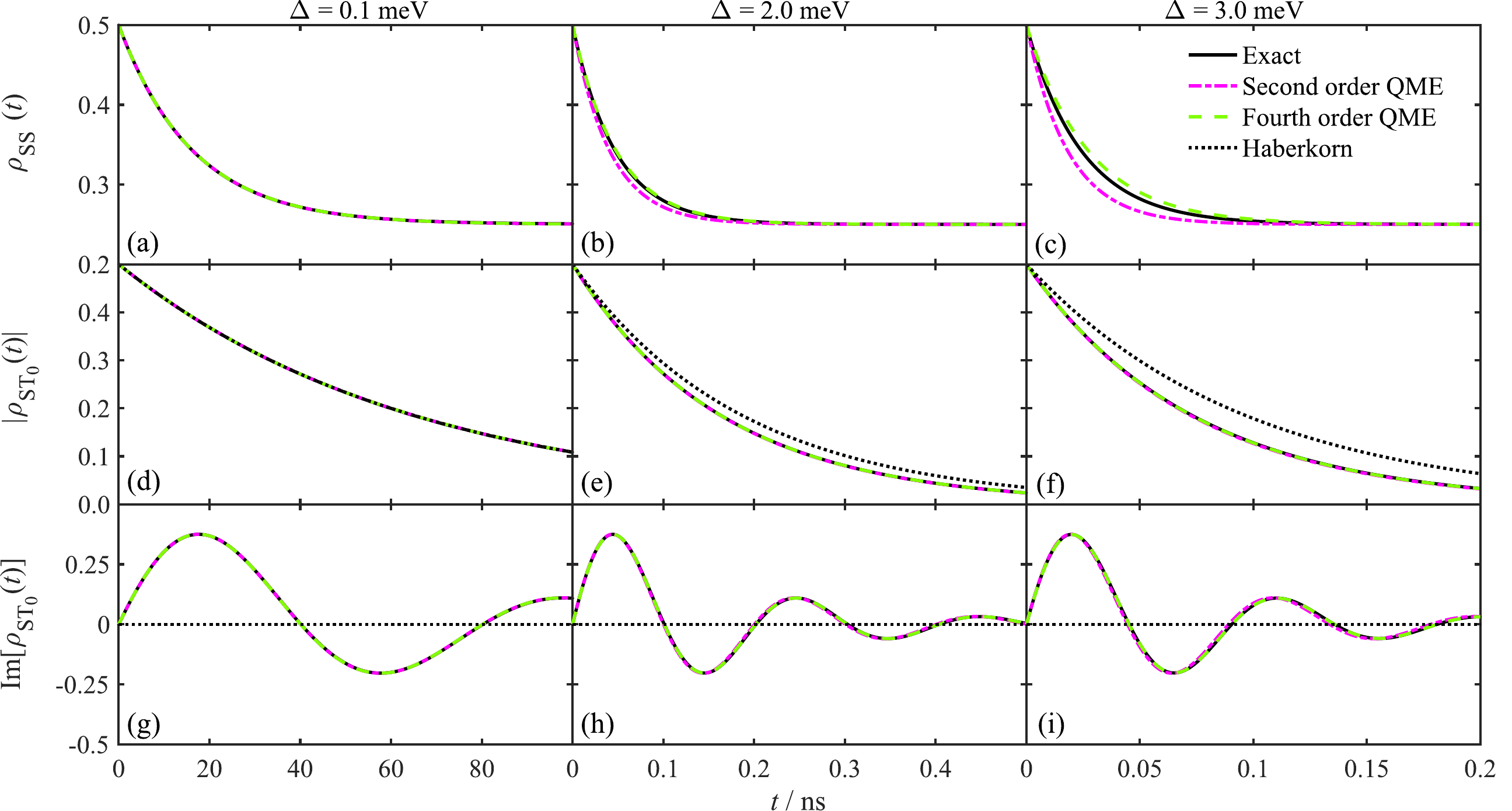}
	\caption{The singlet population, absolute value and imaginary part of the singlet-triplet coherence for the radical pair Model I with three different values of the diabatic coupling calculated using HEOM and the three different master equations. $\Delta = 0.1$ meV for (a), (d) and (g), $\Delta = 2.0$ meV for (b), (e) and (h), and $\Delta = 3.0$ meV for (c), (f) and (i).  }\label{model1-fig}
\end{figure*}
The second and fourth order rate constants, $k_\mathrm{f}$ and $k_\mathrm{b}$, electron spin couplings, $J$, and dephasing rates, $k_\mathrm{d}$, are obtained using the analytic expressions for the spin-boson model which are given in appendices \ref{fourthorder-app} and \ref{spinboson-app}. The parameters are calculated by discretizing the spectral density into a finite set of modes using the standard procedure.\cite{Richardson2015, Wang2003,Berkelbach2012} We find that a discretization into 1000 modes gives converged results for all parameters. The sets of calculated master equation parameters for both models are given in table \ref{parameters-table}.

For Model I the master equations, Eq.~\eqref{trad-me-eqn}, Eq.~\eqref{meq-second-order-eqn} and Eq.~\eqref{meq-fourth-order-eqn}, are analytically soluble. The resulting expressions for the matrix elements of $\rho_{1\sys}(t)$ and $\rho_{2\sys}(t)$ are
\begin{align}
\begin{split}
\rho_{\sing\sing}(t) &= \ev{\rho_{1\sys}(t)}{\sing} =\frac{1}{2} \frac{k_\mathrm{b}+k_\mathrm{f}e^{-(k_\mathrm{f}+k_\mathrm{b})t}}{k_\mathrm{f}+k_{\mathrm{b}}},\label{rhoSS-eqn} \\
\rho_{\sing\trip_0}(t) &=\mel{\sing}{\rho_{1\sys}(t)}{\trip_0}= \frac{1}{2}e^{- 2 i J t/\hbar  - (k_\mathrm{f}/2 + k_\mathrm{d})t},\\
\rho_{\trip_0\sing}(t) &=\mel{\trip_0}{\rho_{1\sys}(t)}{\sing}= \rho_{\sing\trip_0}(t)^* ,\\
\rho_{\trip_0\trip_0}(t) &= \ev{\rho_{1\sys}(t)}{\trip_0} = \frac{1}{2},\\
\rho_{2}(t) &= \ev{\rho_{2\sys}(t)}{\sing} = \frac{1}{2} - \rho_{\sing\sing}(t).
\end{split}
\end{align}
In Model I $k_\mathrm{f}=k_\mathrm{b}$ because in this model $\epsilon = 0$. In the second order master equation $k_\mathrm{d}$ is zero and for the Haberkorn master equation, Eq.~\eqref{trad-me-eqn}, $J = k_\mathrm{d}=0$ (i.e. we still account for the back-reaction in the way predicted by the second order master equation).  For Model II the master equations form a set of linear equations which can be solved numerically using standard techniques. 

\begin{table}
	\begin{tabular}{lcc}
		 & Model I & Model II \\
		 \hline
		 $\hbar^2 k_\mathrm{f}^{(2)}/2\Delta^2$ & $0.66094$ ns& $0.24057$ ns \\
		 $\hbar^2 k_\mathrm{b}^{(2)}/2\Delta^2$ & $0.66094$ ns& $5.0270\times10^{-3}$ ns \\
		 $2\hbar J^{(2)} / \Delta^2 $ & $-3.3952$ ns & $-2.0789$ ns\\
		 $\hbar^4 k_\mathrm{f/b}^{(4)}/2\Delta^4$ & $-1.0634\times 10^{-5}$ ps\textsuperscript{3}& -- \\
		 $2\hbar^3 J^{(4)}/\Delta^4$& $2.4370\times 10^{-6}$ ps\textsuperscript{3}& -- \\
		 $\hbar^4 k_\mathrm{d}^{(4)}/\Delta^4$&$1.0447\times10^{-5}$ ps\textsuperscript{3} & --

	\end{tabular}
\caption{Parameters for the master equations for Models I and II, calculated using expressions given in the main text and Appendices \ref{fourthorder-app} and \ref{spinboson-app}. }\label{parameters-table}
\end{table}

\subsection{Results}

\subsubsection{Model I}

In Fig. \ref{model1-fig} we compare our master equation and exact HEOM simulations for radical pair Model I for a range of values of the diabatic coupling between 0.1 and 3.0 meV. In panels (d)-(i) we also plot the Haberkorn prediction for the time evolution of the singlet-triplet coherence using the numerically exact forward rate constant which is obtained by fitting $\rho_{\sing\sing}(t)$ (panels (a)-(c)) from the HEOM simulation to a function of the form given in Eq.~\eqref{rhoSS-eqn}. 

The radical pair singlet population, $\rho_{\sing\sing}(t)$, shown in panels (a)-(c), is captured qualitatively for all values of $\Delta$ by both the second and fourth order master equations. The Haberkorn fit to $\rho_{\sing\sing}(t)$, using Eq.~\eqref{rhoSS-eqn}, is numerically exact ($R^2 = 1$), which demonstrates the validity of incoherent recombination approximation for this model. The fitted rate constants are $k_\mathrm{f} = 0.03054\text{ ns}^{-1}, 10.63\text{ ns}^{-1},\text{ and } 20.58\text{ ns}^{-1}$ for $\Delta = 0.1, 2.0\text{ and }3.0\text{ meV}$ respectively. As $\Delta$ increases the agreement between the exact results and the master equation results decreases, with the second order quantum master equation (QME) increasingly overestimating the forward and backward rate constants and the fourth order QME increasingly underestimating them. This is of course unsurprising given that our master equations are derived from perturbation theory. Improvements to perturbation theory can be made using the Pad\'e-approximant $ k\simeq k^{(2)}/(1-k^{(4)}/k^{(2)})$ for the rate constant, as has been explored in work by other authors.\cite{Gong2015, Cho1997, Mavros2014,Golosov2001, Sparpaglione1988}

The absolute value of the singlet-triplet coherence, $|\rho_{\sing\trip_0}(t)|$, is shown in panels (d)-(f) of Fig. \ref{model1-fig}. For the master equations, $|\rho_{\sing\trip_0}(t)| = (1/2)e^{-(k_\mathrm{f}/2 + k_\mathrm{d})t}$, which depends only on $k_\mathrm{d}$ and $k_\mathrm{f}$ and not on $J$. Panels (d)-(f) show that the fourth order QME provides as good a description of the evolution of the coherences as it does for $\rho_{\sing\sing}(t)$ in panels (a)-(c). It is at first surprising that the second order QME provides an equally good description of the evolution of  $|\rho_{\sing\trip_0}(t)|$. This is because for the parameters in Model I $k_\mathrm{f}^{(4)}/2 + k_\mathrm{d}^{(4)}\approx 0$ (see Table~I), and therefore the total decay rate of $|\rho_{\sing\trip_0}(t)|$ for both the second and fourth order QMEs is approximately $k_\mathrm{f}^{(2)}$. In other words, because the second order QME overestimates $k_\mathrm{f}$ for larger $\Delta$, it coincidentally describes $|\rho_{\sing\trip_0}(t)|$ very well for all values of $\Delta$ examined here. The Haberkorn prediction for the evolution of $|\rho_{\sing\trip_0}(t)|$, which uses the numerically exact $k_\mathrm{f}$ but which does not include any additional dephasing, increasingly underestimates the decay rate as $\Delta$ increases, which shows that there is in fact additional singlet-triplet dephasing for larger values of the diabatic coupling strength. 

Furthermore, the Haberkorn master equation, which does not include a reactive electron spin coupling term, fails to capture the evolution of the imaginary part of the singlet-triplet coherences, $\Im[\rho_{\sing\trip_0}(t)]$, which is shown in panels (g)-(i) of Fig. \ref{model1-fig}. Our master equations capture the oscillation frequency, which arises due to the reactive scalar electron spin coupling $J$, exceptionally well in these examples. The most significant deviation is for $\Delta=3.0$ meV, shown in panel (i), where the second order master equation slightly overestimates the oscillation frequency. 

This model demonstrates that the conventional Haberkorn reaction operator provides an accurate description of the reduced density operators in the small $\Delta$ limit, provided the second order correction to the scalar electron spin coupling is included. The additional singlet-triplet dephasing term becomes more significant for larger values of $\Delta$. However in the non-adiabatic limit, $\Delta \to 0$, the most important terms are the Haberkorn reaction term and the reactive electron spin coupling, which provide a sufficient description of the dynamics. 

\subsubsection{Model II}

\begin{figure}[t]
	\includegraphics[width=0.475\textwidth]{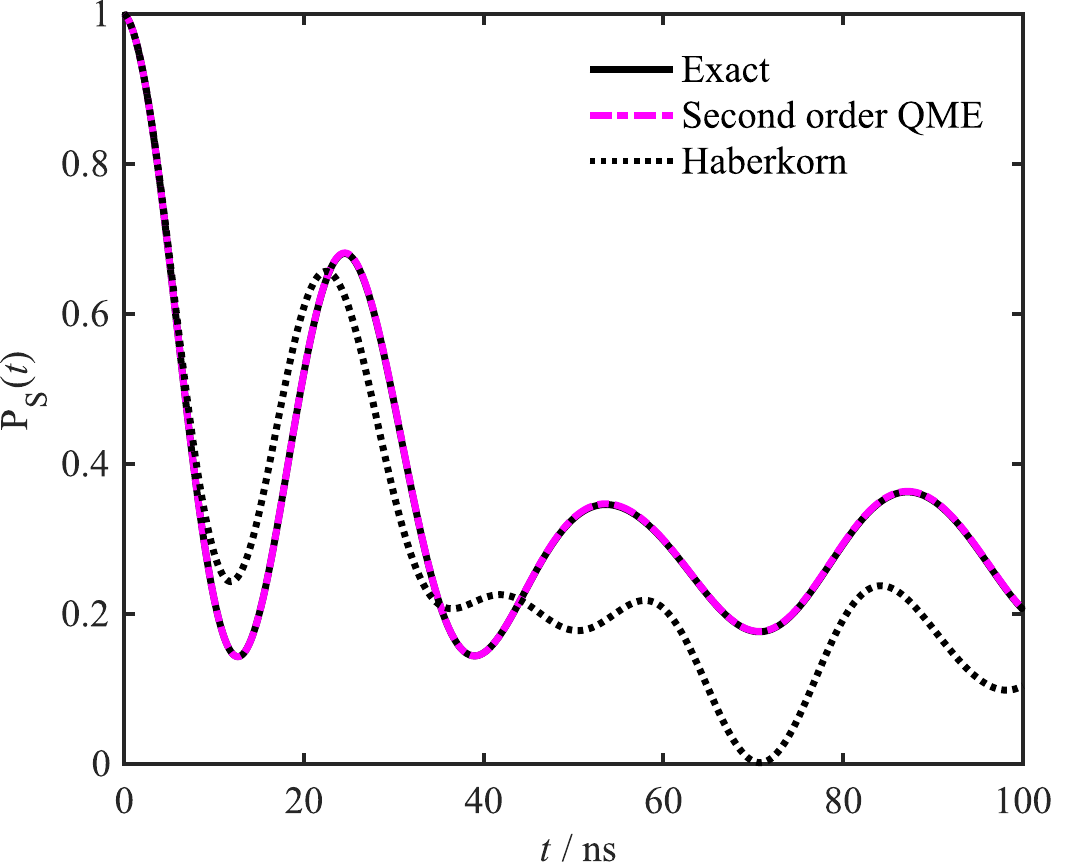}
	\caption{The radical pair singlet population as a function of time for the radical pair Model II. The exact HEOM results, second order quantum master equation (QME) and Haberkorn master equation are shown. The Haberkorn master equation uses the Fermi golden rule rate constant but does not include the reactive electron spin coupling.}\label{model2-fig}
\end{figure}
In Fig. \ref{model2-fig} we plot the radical pair singlet population, $\mathrm{P}_\sing(t) = \Tr_\sys[P_\sing \rho_{1\sys}(t)]$, for Model II, comparing the exact HEOM results, the second order QME results and the Haberkorn QME results. The Haberkorn master equation results use the Fermi golden rule rate $k_\mathrm{f}^{(2)}$ given in Eq.~\eqref{kf-2-eqn}, which appears in the second order QME, but the reactive electron spin coupling term is excluded. Our second order QME quantitatively agrees with the exact HEOM results. However, for times greater than 10 ns the Haberkorn master equation results deviate significantly from the exact results. Both the frequencies and magnitudes of the oscillations in the singlet radical pair population are incorrect for the Haberkorn master equation.

This simple model illustrates two important points. Firstly for this model $H_{1\sys}\neq 0$, but in our results we do not need to account for this in calculating the reaction rate constants for the QME as is demonstrated by the agreement between our second order master equation and the exact results. This illustrates the validity of the field independent rate approximation. Secondly, these results show that for a physically reasonable model for the radical pair electron transfer, it is \textit{essential} to include the reactive electron spin coupling term. Previous derivations of the Haberkorn reaction operator\cite{Ivanov2010} have ignored this term, and because these were not so directly related to reaction rate theory it was difficult to say whether or not this was valid. The present results, combined with our expression relating the Marcus-Hush theory of electron transfer to the reactive electron spin coupling [Eq.~\eqref{marcus-spincoup-eqn}], now provide strong evidence that the reactive contribution the scalar electron spin coupling is not in general negligible.

\section{Concluding Remarks}\label{disc-sec}

In this paper we have used the well-established theory of electron transfer reactions\cite{Golosov2001, Cho1997, Sparpaglione1988, Coalson1994}  to derive quantum master equations for spin selective electron transfers of radical pairs. Our results confirm the validity of the well-known Haberkorn master equation with two corrections. Firstly, we find that there is a reactive contribution to the scalar electron spin coupling that results from the coupling between the spin, electronic and nuclear degrees of freedom -- this arises at leading order in the diabatic coupling along with the Haberkorn reaction term. Secondly, a Jones-Hore-like\cite{Jones2010,Jones2011}  reactive contribution to the singlet-triplet dephasing rate emerges at fourth order in the diabatic coupling. These additional terms are simple and can be straightforwardly included in quantum and semiclassical simulations of radical pairs.\cite{Schulten1978,Manolopoulos2013,Lewis2014,Lawrence2016,Fay2017,Lewis2016,Lindoy2018} 

We have also derived a simple expression for the reactive scalar electron spin coupling within the framework of Marcus theory. This expression, Eq.~\eqref{marcus-spincoup-eqn}, is fully consistent with the Marcus-Hush expression for the singlet and triplet electron transfer rate constants and gives the reactive scalar electron spin coupling solely in terms of the parameters appearing in the Marcus-Hush formula for the rate constants. 

We have validated our master equations, and the approximations required to derive them, by comparison with exact numerical simulations for a simple, but widely studied, model of condensed phase electron transfer. In particular, these results demonstrate the importance of including the reactive contribution to the scalar electron spin coupling. The approach taken in this work can also be generalised to more complex systems with multiple reaction pathways and/or more complex mechanisms of electron transfer, such as long range superexchange and hopping mechanisms.\cite{Jang2013,Wasielewski2006,Hu1989} 

At this point it is worth noting that although the Haberkorn reaction term does not contain electron spin coupling or singlet-triplet dephasing terms, these terms often appear anyway in models of radical pair reactions because there are other well-known physical mechanisms that give rise to them. Direct exchange and superexchange interactions between radicals give rise electron spin coupling,\cite{Wasielewski2006, Steiner1989} and modulation of these interactions by molecular motion leads to singlet-triplet dephasing.\cite{Lawrence2016, Kattnig2016, Breuer2001, Shushin1991} The coupling strength and dephasing rate are often fitted to experimental data, such as Magnetically Affected Reaction Yield (MARY) spectra\cite{Rodgers2009,Steiner1989} or Time Resolved Electron Paramagnetic Resonance (TREPR) spectra,\cite{Wasielewski2006, Bittl2005} and these fitted parameters would naturally contain both reactive and non-reactive contributions. As previously noted by Jones and Hore in the context of the dephasing rate,\cite{Jones2010,Jones2011} disentangling these contributions in real experimental data is likely to be very difficult because of the difficulty in quantitatively predicting any single contribution to the dephasing rate or electron spin coupling. 

Despite the difficulties in separating reactive and non-reactive contributions to the electron spin coupling, we would like to suggest two simple experiments that might be able to disentangle them. Suppose that for a given radical pair the singlet recombination rate is much larger than the triplet recombination rate and that the dominant contribution to the reactive electron spin coupling comes from the singlet pathway ($\Delta_\sing\gg\Delta_\trip$). Using the Marcus-Hush theory expressions for $k_\sing\equiv k_{\rm f,cl}^{(2)}$ and $J\equiv J_{\rm cl}^{(2)}$  in Eqs.~\eqref{marcus-rate-eqn} and \eqref{marcus-spincoup-eqn}, we see that their ratio depends only on the Marcus-Hush activation energy $E_\mathrm{a} = (\lambda-\epsilon)^2/4\lambda$ and the sign of $\epsilon-\lambda$,
\begin{align}
\frac{2 J }{\hbar k_\sing} = \frac{1}{2}\mathrm{sign}(\epsilon-\lambda) \mathrm{erfi}\left(\sqrt{\frac{E_\mathrm{a}}{k_\mathrm{B}T}}\right).
\end{align}
The recombination rate constants and total scalar electron spin coupling can be measured by TREPR or MARY spectroscopy and $E_\mathrm{a}$ can be determined from the temperature dependence of $k_\sing$. By comparing the theoretical ratio of $2J/\hbar$ to $k_\sing$ one could determine the relative sizes of the reactive and non-reactive contributions to the total electron spin coupling. This argument applies equally in the case where the triplet rate is much larger than the singlet, but the sign of $2J/\hbar k_\trip$ is reversed.

An alternative suggestion is to measure the sign of the electron spin coupling of a molecular radical ion pair by TREPR spectroscopy\cite{Dance2006} under conditions where it is known that either the singlet or triplet reaction pathway dominates. Some radical pair systems undergo an inversion of the dominant reaction pathway on changing solvent conditions.\cite{Weiss2004a} This would not be expected to change the sign of any non-reactive exchange coupling, which is typically controlled by through-bond interactions, but would invert the sign of the reactive spin-coupling (see appendix \ref{trippath-app}). 

The existence of reactive contributions to singlet-triplet dephasing rates and electron spin coupling in electron transfer reactions in radical pairs may have significant implications for theoretical investigations into magnetoreception in birds and other animals. In many models of the radical pair based avian magnetoreceptor, electron spin coupling is neglected on the assumption that the two radicals are well-separated in space.\cite{Hiscock2016,Worster2016} We now however have a strong theoretical basis for saying that an electron spin coupling will be present due to the spin-selective radical pair recombination.

Overall, we hope that this work will put the Haberkorn master equation, and all previous studies of electron transfer reactions in radical pairs which have used it, on a stronger theoretical footing. In the non-adiabatic limit the Haberkorn reaction term gives the correct description of spin-selective electron transfer processes, provided a reactive scalar electron spin coupling is also included in the master equation. Other master equations proposed by other authors do not correctly describe the spin-selective recombination process for the type of reaction considered in this work. 

Finally, we should note that a reactive contribution to the scalar electron spin coupling, similar to that proposed here, has been suggested previously by Vitalis and Kominis in Ref. \onlinecite{Vitalis2014}. However, their description of the recombination process is quite different to ours, and it is not related in such a direct way to standard electron transfer rate theory. Moreover we have explicitly verified the accuracy of the scalar electron spin coupling terms in our second and fourth order master equations by comparison with exact HEOM results in physically reasonable electron transfer regimes.

\section*{Supplementary Material}

In the supplementary material we outline the generalisation of our fourth order quantum master equation, Eq. \eqref{meq-fourth-order-eqn}, to the case where a triplet recombination pathway is included, and to the case where the diabatic coupling is non-constant, i.e. $f_\sing \neq 1$ in Eq. \eqref{hamiltonian-eqn}. We demonstrate that the general form of our fourth order master equation is unchanged in these cases, but the expressions for the rate, dephasing and spin coupling constants appearing in the master equation are changed.

\begin{acknowledgements}
	
We would like to thank Joseph Lawrence for many very helpful discussions. We are also thankful to Peter Hore for his comments on the first draft of this manuscript. Thomas Fay is supported by a Clarendon Scholarship from Oxford University, an E.A. Haigh Scholarship from Corpus Christi College, Oxford, and by the EPRSC Centre for Doctoral Training in Theory and Modelling in the Chemical Sciences, EPSRC Grant No. EP/L015722/1. Lachlan Lindoy is supported by a Perkin Research Studentship from Magdalen College, Oxford, an Eleanor Sophia Wood Postgraduate Research Travelling Scholarship from the University of Sydney, and by a James Fairfax Oxford Australia Scholarship.
	
\end{acknowledgements}

\appendix

\section{Proof of Equation \eqref{projpropb-eqn}}\label{projprop-app}
In order to prove the result in Eq.~\eqref{projpropb-eqn}, we first note that $V^{\ipi}(t)$ is given by
\begin{align}
V^{\ipi}(t) = \Delta\left(G(t)e^{iH_{1\sys}t/\hbar}P_\sing\dyad{1}{2} + \dyad{2}{1}P_\sing e^{-iH_{1\sys}t/\hbar}G(t)^\dag \right),
\end{align}
where
\begin{align}
G(t) = e^{+iH_{1\nuc}t/\hbar }e^{-iH_{2\nuc}t/\hbar }.\label{Gdef-eqn}
\end{align}
We also note that $G(t)e^{iH_{1\sys}t/\hbar}P_\sing$ is an operator that does not act on the electronic state degree of freedom, and that for any operator $A$, $\pP A$ is diagonal in the diabatic electronic state basis. I.e., $\pP A = \dyad{1}B + \dyad{2}C$, where $B$ and $C$ are operators that only act on the nuclear and spin degrees of freedom. 

First let us consider $\pL_V^\ipi(t_1)\pP A$. From the definition of $\pL_V^\ipi(t)$ in Eq.~\eqref{ipiliouvillian-eqn}, it is clear that this will be of the form 
\begin{align}
\pL_V^\ipi(t_1)\pP A = \dyad{1}{2}B'+\dyad{2}{1}C',
\end{align}
where $B'$ and $C'$ are operators on the nuclear and spin degrees of freedom. $\pP$ removes any off-diagonal terms in the diabatic electronic state basis, $\pP(\dyad{1}{2}B')=0$ and $\pP(\dyad{2}{1}C')=0$, and therefore the result in Eq.~\eqref{projpropb-eqn} clearly holds for $n=0$,
\begin{align}
\pP\pL_V^\ipi(t_1)\pP = 0.
\end{align}

It is now straightforward to extend this to all products of an odd number of $\pL_V^\ipi(t_k)$s. Again from the definition of $\pL_V^\ipi(t)$ in Eq.~\eqref{ipiliouvillian-eqn}, we see that 
\begin{align}
\pL_V^\ipi(t_2)\pL_V^\ipi(t_1)\pP A = \dyad{1}{1}B''+\dyad{2}{2}C'',
\end{align}
where $B''$ and $C''$ again only operate on the nuclear and spin degrees of freedom. We see that this is of the same form as $\pP A$ and therefore
\begin{align}
\pP\pL_V^\ipi(t_3)\pL_V^\ipi(t_2)\pL_V^\ipi(t_1)\pP = 0.
\end{align}
Iterating this argument, we see that in general an even number of $\pL_V^\ipi(t_k)$s acting on $\pP A$ gives an operator that only contains terms which are diagonal in the diabatic electronic state basis, and an odd number of $\pL_V^\ipi(t_k)$s acting on $\pP A$ gives an operator that only contains terms that are off-diagonal in this basis. Because $\pP$ removes any terms which are off-diagonal, Eq.~\eqref{projpropb-eqn} clearly holds for any product of an odd number of $\pL_V^\ipi(t_k)$s.\\
\section{Including the Triplet Reaction Pathway}\label{trippath-app}
The same techniques outlined in the main text can be used to derived second order master equations for a radical pair undergoing both singlet and triplet state selective electron transfer reactions, i.e. when we consider the full Hamiltonian Eq.~\eqref{hamiltonian-eqn} with $\Delta_\trip\neq 0$. In doing so we assume there are initially no coherences between the singlet and triplet product states, and that initially the nuclei on each state are at thermal equilibrium on that diabatic potential energy surface. The general form of the second order master equations for the radical pair spin density operator $\rho_{1\sys}(t)$, and the singlet and triplet product spin density operators, $\rho_{2\sys}^{\sing}(t)$ and $\rho_{2\sys}^{\trip}(t)$, is
\begin{widetext}
	\begin{subequations}\label{meq-second-order-full-eqn}
		\begin{gather}
		\dv{t}\rho_{1\sys}(t) = -\frac{i}{\hbar}\left[H_{1\sys},\rho_{1\sys}(t)\right] -\frac{i}{\hbar}\left[(-2J^{(2)})\vb{S}_1\cdot\vb{S}_2,\rho_{1\sys}(t)\right]  - \left\{ K_{\rm s}^{(2)}, \rho_{1\sys}(t) \right\} + k_\mathrm{b,\sing}^{(2)} P_\sing \rho_{2\sys}^\sing(t) P_\sing+k_\mathrm{b,\trip}^{(2)} P_\trip \rho_{2\sys}^\trip(t) P_\trip, \\
		\dv{t}\rho_{2\sys}^\sing(t) = k_\mathrm{f,\sing}^{(2)}P_\sing \rho_{1\sys}(t)P_\sing - k_\mathrm{b,\sing}^{(2)} \rho_{2\sys}^\sing(t), \\
		\dv{t}\rho_{2\sys}^\trip(t) = k_\mathrm{f,\trip}^{(2)}P_\trip \rho_{1\sys}(t)P_\trip - k_\mathrm{b,\trip}^{(2)} \rho_{2\sys}^\trip(t).
		\end{gather}
	\end{subequations}
\end{widetext}
Here $K_{\rm s}^{(2)}$ is the Haberkorn reaction operator with the second order forward rate constants for the two spin selective recombination pathways,
\begin{align}
K_{\rm s}^{(2)} = \frac{k_{\mathrm{f},\sing}^{(2)}}{2}P_\sing + \frac{k_{\mathrm{f,\trip}}^{(2)}}{2}P_\trip,
\end{align}
and $J^{(2)}$ is now the full reactive electron spin coupling, which is a difference of singlet and triplet components,
\begin{align}
J^{(2)} = J_\sing^{(2)} - J_\trip^{(2)}.
\end{align}
The triplet contribution appears with the opposite sign to the singlet contribution because $P_\sing = \frac{1}{4} - \vb{S}_1\cdot\vb{S}_2$ and $P_\trip = \frac{3}{4} + \vb{S}_1\cdot\vb{S}_2$. The expressions for the parameters in these equations are the same as the expressions appearing in the main text but with $H_{2\nuc}$ replaced with $H_{2\nuc}^\sing$ for $k_\mathrm{f,\sing}^{(2)}$, $k_\mathrm{b,\sing}^{(2)}$ and $J_\sing^{(2)}$, and with $H_{2\nuc}^\trip$ for $k_\mathrm{f,\trip}^{(2)}$, $k_\mathrm{b,\trip}^{(2)}$ and $J_\trip^{(2)}$. 

The extension of this to fourth and higher orders in $\Delta_\sing$ and $\Delta_\trip$ is presented in the Supplementary Information. The fourth order master equation is of the same form as Eq.~\eqref{meq-second-order-full-eqn} but with an additional singlet-triplet dephasing term in the equation for $\rho_{1\sys}(t)$. It should also be noted that when the Condon approximation is \textit{not} made, i.e. when $f_\sing$ and $f_\trip$ in Eq.~\eqref{hamiltonian-eqn} are \textit{not} assumed to be identity operators, the form of the master equation is not changed, but the expressions for $k_\mathrm{f}$, $k_\mathrm{b}$, $J$ and $k_\mathrm{d}$ are modified. This is also discussed in the Supplementary Information.
\section{Fourth Order Rate Expressions}\label{fourthorder-app}
The expressions for the fourth order contributions to the reaction rate constants, singlet-triplet dephasing rate and scalar electron spin coupling are related to the correlation functions $c_{1}^{(2)}(t)$ and $c_{2}^{(2)}(t)$, defined in Eq.~\eqref{corrfunc1-eqn} and Eq.~\eqref{corrfunc2-eqn}, as well as the three-time correlation functions defined below
\begin{align}
c_{1}^{(2)}(t_0,t_1,t_2) &= \Tr_\nuc\left[\rho_{1\nuc}^\eq G(t_0)G(t_1)^\dag G(t_2)\right], \\
c_{2}^{(2)}(t_0,t_1,t_2) &= \Tr_\nuc\left[\rho_{2\nuc}^\eq G(t_0)^\dag G(t_1) G(t_2)^\dag\right],
\end{align}
where $G(t)$ is given by Eq.~\eqref{Gdef-eqn}. 
The fourth order contribution to the rate constant is\cite{Golosov2001} 
\begin{widetext}
	\begin{align}\label{fourth-order-rate-eqn}
	\begin{split}
	k_\mathrm{f}^{(4)} = -\frac{2\Delta^4}{\hbar^4}\int_0^\infty\dd{t_0}\int_0^{t_0}\dd{t_1}\int_0^{t_1}\dd{t_2}\bigg(&\Re\left[c_{1}^{(4)}(t_0,t_1,t_2)+c_{1}^{(4)}(t_2,t_1,t_0)+c_{1}^{(4)}(t_1,t_0,t_2)+c_{1}^{(4)}(t_2,t_0,t_1)\right] \\
	&- 2\Re\left[c_1^{(2)}(t_0-t_1)\right]\Re\left[c_1^{(2)}(t_2)\right]- 2\Re\left[c_2^{(2)}(t_0-t_1)\right]\Re\left[c_1^{(2)}(t_2)\right]\bigg).
	\end{split}
	\end{align}
	The fourth order contribution to the back-reaction rate, $k_{\mathrm{b}}^{(4)}$, is obtained by simply swapping the state indices 1 and 2 in the above expression. The fourth order contribution to the electron spin coupling is
	\begin{align}
	\begin{split}
	J^{(4)} = -\frac{\Delta^4}{2\hbar^3}\int_0^\infty\dd{t_0}\int_0^{t_0}\dd{t_1}\int_0^{t_1}\dd{t_2}\bigg(&\Im\left[c_{1}^{(4)}(t_0,t_1,t_2)\right]-\Im\left[c_1^{(2)}(t_0-t_1)c_1^{(2)}(t_2)\right]\bigg).
	\end{split}
	\end{align}
	Finally, the fourth order singlet-triplet dephasing rate constant is
	\begin{align}
	\begin{split}
	k_\mathrm{d}^{(4)} = \frac{\Delta^4}{\hbar^4}\int_0^\infty\dd{t_0}\int_0^{t_0}\dd{t_1}\int_0^{t_1}\dd{t_2}\bigg(&\Re\left[c_{1}^{(4)}(t_2,t_1,t_0)+c_{1}^{(4)}(t_1,t_0,t_2)+c_{1}^{(4)}(t_2,t_0,t_1)\right] \\
	&- \Re\left[c_1^{(2)}(t_0-t_1)^*c_1^{(2)}(t_2)\right]- 2\Re\left[c_2^{(2)}(t_0-t_1)\right]\Re\left[c_1^{(2)}(t_2)\right]\bigg).
	\end{split}
	\end{align}
\end{widetext}
The rate constant expression in Eq.~\eqref{fourth-order-rate-eqn} is consistent with that obtained previously by others -- e.g., Golosov and Reichman in Ref. \onlinecite{Golosov2001}. The triple integrals in these expressions were evaluated numerically to calculate the parameters for the fourth order master equation results for Model I in Fig. \ref{model1-fig}.
\section{Spin Boson Correlation Functions}\label{spinboson-app}
Analytic expressions for the spin-boson correlation functions can be obtained using harmonic oscillator coherent states.\cite{Mavros2014} These expressions are
\begin{align}
c_1^{(2)}(t) &= \exp(\zeta(t)+i\epsilon t /\hbar), \\
\begin{split}
c_{1}^{(4)}(t_0,t_1,t_2)  &= \exp\big(\zeta(t_1) + \zeta(t_0-t_2) \\
&\ - \zeta(t_0) - \zeta(t_2) - \zeta(t_0-t_1) \\
&\ - \zeta(t_1-t_2)+i\epsilon (t_0+t_2) /\hbar\big).
\end{split}
\end{align}
where the function $\zeta(t)$ is given by
\begin{align}
\zeta(t) &= -\kappa(t)+i\phi(t) .
\end{align} 
in which $\kappa(t)$ and $\phi(t)$ are related to the spectral density by
\begin{align}
\kappa(t) &= \frac{4}{\pi \hbar } \int_0^\infty\dd{\omega} \frac{J(\omega)}{\omega^2}\coth(\frac{\beta\hbar\omega}{2})(1-\cos(\omega t)) \\
\phi(t) &= -\frac{4}{\pi \hbar }\int_0^\infty\dd{\omega} \frac{J(\omega)}{\omega^2}\sin(\omega t) 
\end{align}
The equivalent expressions for $c_{2}^{(2)}(t)$ and $c_{2}^{(4)}(t_0,t_1,t_2)$ are obtained by changing $\epsilon$ in the above expressions to $-\epsilon$. These expressions give rate constants consistent with those obtained previously.\cite{Mavros2014} We numerically integrated these functions to calculate the rate constants, dephasing rates and scalar electron spin coupling strengths appearing in the QMEs for Models I and II.

\bibliography{haberkorn-paper}

\end{document}



\title{Supplementary Information \\Spin-selective electron transfer reactions of radical pairs: beyond the Haberkorn master equation}
\author[1]{Thomas P. Fay}
\author[1]{Lachlan P. Lindoy}
\author[1]{David E. Manolopoulos}
\affil[1]{\em Department of Chemistry, University of Oxford, Physical and Theoretical Chemistry Laboratory, South Parks Road, Oxford, OX1 3QZ, UK}
\date{}
\maketitle

\section*{Fourth order master equation including the triplet pathway}

The fourth order master equations for a radical pair system with a triplet recombination pathway are obtained using the techniques outlined in the main text. We assume there are initially no coherences between singlet and triplet product states, and we modify the Liouville space projection superoperator as follows,
\begin{align}
\pP_2 W(t) = \dyad{2}{2}\frac{e^{-\beta H_{2\nuc}^\sing}}{\Tr_\nuc[e^{-\beta H_{2\nuc}^\sing}]}P_\sing\Tr_\nuc[\ev{W}{2}]P_\sing+\dyad{2}{2}\frac{e^{-\beta H_{2\nuc}^\trip}}{\Tr_\nuc[e^{-\beta H_{2\nuc}^\trip}]}P_\trip\Tr_\nuc[\ev{W}{2}]P_\trip.
\end{align}
We describe the system with spin density operators for the singlet and triplet products
\begin{align*}
\rho_{2\sys}^\sing(t) &= P_\sing\Tr_\nuc\left[\ev{\pP_2 W(t)}{2}\right]P_\sing, \\
\rho_{2\sys}^\trip(t) &= P_\trip\Tr_\nuc\left[\ev{\pP_2 W(t)}{2}\right]P_\trip.
\end{align*}
With this established, the fourth order master equations for the spin density operators including the triplet pathway are
	\begin{subequations}\label{meq-fourth-order-full-eqn}
		\begin{gather}
		\begin{split}
		\dv{t}\rho_{1\sys}(t) = -&\frac{i}{\hbar}\left[H_{1\sys},\rho_{1\sys}(t)\right] -\frac{i}{\hbar}\left[(-2J)\vb{S}_1\cdot\vb{S}_2,\rho_{1\sys}(t)\right]  - \left\{ K_{\rm s}, \rho_{1\sys}(t) \right\} \\
		-\ &k_\mathrm{d}(P_\sing \rho_{1\sys}(t) P_\trip + P_\trip \rho_{1\sys}(t) P_\sing)+ k_\mathrm{b,\sing} P_\sing \rho_{2\sys}^\sing(t) P_\sing+k_\mathrm{b,\trip} P_\trip \rho_{2\sys}^\trip(t) P_\trip, 
		\end{split}\\
		\dv{t}\rho_{2\sys}^\sing(t) = k_\mathrm{f,\sing}^{(2)}P_\sing \rho_{1\sys}(t)P_\sing - k_\mathrm{b,\sing}^{(2)} \rho_{2\sys}^\sing(t), \\
		\dv{t}\rho_{2\sys}^\trip(t) = k_\mathrm{f,\trip}^{(2)}P_\trip \rho_{1\sys}(t)P_\trip - k_\mathrm{b,\trip}^{(2)} \rho_{2\sys}^\trip(t).
		\end{gather}
	\end{subequations}
The expressions for the forward and backward spin selective rate constants are obtained using the expressions for $k^{(4)}_\mathrm{f}$ and $k^{(4)}_\mathrm{b}$ in appendix C in the main text. The singlet rate constants are obtained by replacing with $H_{2\nuc}$ with $H_{2\nuc}^\sing$ and $\Delta$ with $\Delta_\sing$, and the triplet rate constants are obtained by replacing $H_{2\nuc}$ with $H_{2\nuc}^\trip$ and $\Delta$ with $\Delta_\trip$. The dephasing rate $k_\mathrm{d}$ is given by
\begin{align}
k_\mathrm{d} = k_\mathrm{d,\sing}^{(4)} + k_\mathrm{d, \trip}^{(4)} - k_\mathrm{d, \sing\trip}^{(4)}.
\end{align}
It is a sum of the singlet and triplet contributions, given by the expression in appendix C with $H_{2\nuc}$ and $\Delta$ changed appropriately, and a cross term which appears with the opposite sign. This cross term is
	\begin{align}
	\begin{split}
	k_\mathrm{d,\sing\trip}^{(4)} = \frac{\Delta^2_\sing\Delta_\trip^2}{\hbar^4}\int_0^\infty\dd{t_0}\int_0^{t_0}&\dd{t_1}\int_0^{t_1}\dd{t_2}\bigg(\Re\left[c_{1,\sing\trip}^{(4)}(t_2,t_1,t_0)+c_{1,\sing\trip}^{(4)}(t_1,t_0,t_2)+c_{1,\sing\trip}^{(4)}(t_2,t_0,t_1)\right] \\
	&+ \Re\left[c_{1,\trip\sing}^{(4)}(t_2,t_1,t_0)+c_{1,\trip\sing}^{(4)}(t_1,t_0,t_2)+c_{1,\trip\sing}^{(4)}(t_2,t_0,t_1)\right]\\
	&- \Re\left[c_{1,\trip}^{(2)}(t_0-t_1)^*c_{1,\sing}^{(2)}(t_2)\right] - \Re\left[c_{1,\sing}^{(2)}(t_0-t_1)^*c_{1,\trip}^{(2)}(t_2)\right] \bigg).
	\end{split}
	\end{align}
$c_{j,\sing}^{(2)}(t)$ and $c_{j,\trip}^{(2)}(t)$ are defined in the same way as $c_{j}^{(2)}(t)$ but with $H_{2\nuc}$ replaced with $H_{2\nuc}^\sing$ and $H_{2\nuc}^\trip$ respectively. $c_{1,\sing\trip}(t_0,t_1,t_2)$ is defined as
\begin{align}
c_{1,\sing\trip}(t_0,t_1,t_2) = \Tr_\nuc\left[\rho_{1\nuc}^\eq G_\sing(t_0)G_\sing(t_1)^\dag G_\trip(t_2)\right],
\end{align}
where $G_\sing(t)$ and $G_\trip(t)$ are defined the same way as $G(t)$ in appendix A but with the singlet or triplet product nuclear Hamiltonians replacing $H_{2\nuc}$. For $c_{1,\trip\sing}(t_0,t_1,t_2)$ we simply swap the labels $\sing$ and $\trip$. Similarly there is a cross term contributing to the reactive electron spin coupling at fourth order
\begin{align}
J = J_\sing^{(2)} + J_\sing^{(4)} - J_\trip^{(2)} - J_\trip^{(4)} + J_{\sing \trip}^{(4)}.
\end{align}
This term is given by
	\begin{align}
	\begin{split}
	J_\mathrm{\sing\trip}^{(4)} = \frac{\Delta^2_\sing\Delta_\trip^2}{2\hbar^3}\int_0^\infty\dd{t_0}\int_0^{t_0}&\dd{t_1}\int_0^{t_1}\dd{t_2}\bigg(\Im\left[c_{1,\sing\trip}^{(4)}(t_2,t_1,t_0)+c_{1,\sing\trip}^{(4)}(t_1,t_0,t_2)+c_{1,\sing\trip}^{(4)}(t_2,t_0,t_1)\right] \\
	&- \Im\left[c_{1,\trip\sing}^{(4)}(t_2,t_1,t_0)+c_{1,\trip\sing}^{(4)}(t_1,t_0,t_2)+c_{1,\trip\sing}^{(4)}(t_2,t_0,t_1)\right]\\
	&- \Im\left[c_{1,\trip}^{(2)}(t_0-t_1)^* c_{1,\sing}^{(2)}(t_2)\right]+ \Im\left[c_{1,\sing}^{(2)}(t_0-t_1)^* c_{1,\trip}^{(2)}(t_2)\right]\bigg).
	\end{split}
	\end{align}

It should be noted that when we do not make the Condon approximation, in which we assume that $f_\sing$ and $f_\trip$ in the full Hamiltonian $H$ in Eq. (4) are identity operators, the above expressions for the rate constants, dephasing rate and scalar electron spin coupling strength must be modified. However, the form of the master equations remains the same, because the spin state projection operators $P_\sing$ and $P_\trip$ commute with any function of the nuclear coordinates. The modified expressions for the master equation parameters when $f_\sing\neq 1$ and $f_\trip\neq 1$ are obtained by modifying the expressions for the correlation functions. For example, we redefine $c_{1,\sing}^{(2)}(t)$ as 
\begin{align}
c_{1,\sing}^{(2)}(t) = \Tr_\nuc\left[\rho_{1\nuc}^\eq e^{+iH_{1\nuc}t/\hbar}f_\sing e^{-iH_{2\nuc}^\sing t/\hbar}f_\sing^\dag\right],
\end{align}
and we similarly redefine $c_{1,\sing}^{(4)}(t_0,t_1,t_2)$ as
\begin{align}
c_{1,\sing}^{(4)}(t_0,t_1,t_2) = \Tr_\nuc\left[\rho_{1\nuc}^\eq e^{+iH_{1\nuc}t_0/\hbar}f_\sing e^{-iH_{2\nuc}^\sing t_0/\hbar}e^{+iH_{2\nuc}^\sing t_1/\hbar}f_\sing^\dag e^{-iH_{1\nuc} t_1/\hbar}e^{+iH_{1\nuc}t_2/\hbar}f_\sing e^{-iH_{2\nuc}^\sing t_2/\hbar}f_\sing^\dag\right].
\end{align}
Analogous modifications must be made to all other correlation functions when $f_\sing\neq 1$ and $f_\trip\neq 1$.